\documentclass{jfm}
\usepackage{graphicx}
\usepackage{color}
\usepackage{amssymb}
\usepackage{amsmath}
\usepackage{float}

\newcommand{\bB}{\textit{\textbf{B}}}
\newcommand{\bU}{\textit{\textbf{U}}}

\newcommand{\be}{\textit{\textbf{e}}}
\newcommand{\bE}{\textit{\textbf{E}}}

\newcommand{\bH}{\textit{\textbf{H}}}
\newcommand{\bJ}{\textit{\textbf{J}}}

\newcommand{\bq}{\textit{\textbf{q}}}

\newcommand{\ii}{\mathrm{i}}

\newcommand\Rm{\textit{Rm}}
\newcommand{\sign}{\text{sign}}

\shorttitle{Fast and Furious anisotropic dynamo}
\shortauthor{F. Plunian, T. Alboussi\`ere}

\title{Fast and Furious dynamo action in the anisotropic dynamo}

\author{Franck Plunian\aff{1}
  \corresp{\email{Franck.Plunian@univ-grenoble-alpes.fr}},
 \and 
 Thierry Alboussi\`ere\aff{2}}

\affiliation{\aff{1}Univ. Grenoble Alpes, Univ. Savoie Mont Blanc, CNRS, IRD, Univ. Gustave Eiffel, ISTerre, 38000 Grenoble, France
\aff{2}Universit\'e de Lyon, ENSL, UCBL, CNRS, LGL-TPE, 46 all\'ee d'Italie, F-69364 Lyon, France}

\begin{document}

\maketitle

\begin{abstract}
In the limit of large magnetic Reynolds numbers, it is shown that a smooth differential rotation can lead to fast dynamo action, provided that the electrical conductivity or magnetic permeability is anisotropic. 
If the shear is infinite, for example between two rotating solid bodies, the anisotropic dynamo becomes furious, meaning that the magnetic growth rate increases toward infinity with an increasing magnetic Reynolds number.
\end{abstract}


\section{Introduction}
Dynamo action is a magnetic instability that converts part of the kinetic energy of a moving material into magnetic energy, without the aid of a magnet, but provided of course that the material is electrically conducting \citep{Rincon2019,Tobias2021}.
One of the simplest kinematic dynamos is the anisotropic dynamo, which relies on the anisotropy of electromagnetic properties, and for which an exponentially growing magnetic field can be generated by a velocity field as simple as a shear \citep{Ruderman1984, Lortz1989}.
Anisotropic electrical conductivity means that the electric current density $\bJ$ is no longer parallel to the electric field $\bE$, even in the absence of a velocity field $\bU$. Similarly, an anisotropic magnetic permeability means that the magnetic field $\bH$ and the induction field $\bB$ are no longer parallel. In natural objects, anisotropy in electric conductivity may result, in the Earth's core from the anisotropic crystallisation of the inner core \citep{Deuss2014, Ohta2018}, in plasmas from the presence of an external magnetic field \citep{Braginskii1965}, and in spiral galaxies \citep{Brandenburg2005} from their spiral geometries. In contrast, anisotropy of magnetic permeability seems unlikely, at least at large scale. 
However, at the laboratory scale, among the few experiments that succeeded in reproducing a dynamo effect, one involved soft iron \citep{Lowes1963,Lowes1968}, while another worked only in the presence of soft iron propellers \citep{Miralles2013,Kreuzahler2017,Nore2018}, highlighting the crucial role that magnetic permeability can play.
In this paper, the medium is taken as homogeneous, which excludes any source of dynamo action based on spatial variations of electrical conductivity or magnetic permeability \citep{Petrelis2016,Marcotte2021,Gallet2012,Gallet2013}.

The anisotropic dynamo has several features making it unique. 
\begin{itemize}
\item[$(i)$] Defeating Cowling's antidynamo theorem \citep{Cowling1934}, it was shown that fully axisymmetric dynamo action is possible in cylindrical geometry \citep{Plunian2020}. The counterpart in Cartesian geometry makes dynamo possible for two-dimensional plane motion \citep{Ruderman1984,Alboussiere2020}, defeating Zel'dovich's antidynamo theorem \citep{Zeldovich1957}. This reduces the validity of these two antidynamo theorems to the case of isotropic magnetic diffusivity, which was in fact implicitely assumed by Cowling and Zel’dovich. 
\item[$(ii)$] For a sliding motion corresponding to infinite shear, in Cartesian geometry the opposite motions of two superimposed plates \citep{Alboussiere2020}, in cylindrical geometry the opposite rotations of two coaxial cylinders \citep{Plunian2020}, an exact dynamo threshold can be explicitely derived. Furthermore, it was found that the dynamo threshold is small enough to be experimentally tested. 
\item[$(iii)$] The effect on dynamo action of the anisotropy of the magnetic permeability is opposite to that of the electrical conductivity, which even makes the dynamo impossible if the two anisotropies are identical \citep{Plunian2021}. 
\item[$(iv)$] In Cartesian geometry the anisotropic dynamo is found to be fast if the shear is smooth \citep{Ruderman1984}, and furious if the shear is infinite \citep{Alboussiere2020}, the meaning of these two types of dynamo action, fast and furious, being explained below. In cylindrical geometry, there is a priori no reason why it should be different. However this remains to be proven, which is the subject of this paper, for the two cases, a smooth differential rotation and an infinite shear.
\end{itemize}

\begin{figure}
\begin{center}
\includegraphics[scale=0.4]{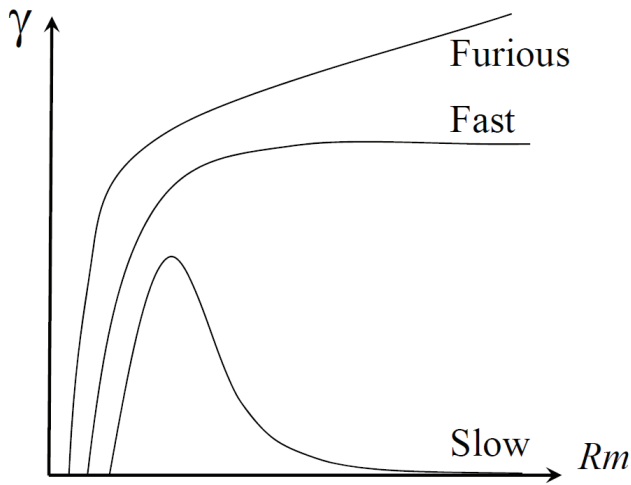}
\caption{Illustration of possible growth rates $\gamma$ versus the magnetic Reynolds number $Rm$, for slow, fast and furious dynamos.}
\label{fig:illustration}
\end{center}
\end{figure}

A kinematic dynamo is said to be fast if, in the limit of large magnetic Reynolds numbers, the magnetic growth rate tends towards, or oscillates around, a positive limit. In this case, the magnetic energy grows on a time-scale smaller than that of magnetic diffusion, typically the advective time scale.
In contrast, a kinematic dynamo is said to be slow if, in the limit of large magnetic Reynolds numbers, the magnetic growth rate tends towards zero, meaning that the dynamo occurs on the magnetic diffusion time scale, or on a time scale between that of advection and magnetic diffusion. This distinction between slow and fast dynamo was first made by \citet{Vainshtein1972} for astrophysical objects like the Sun where the magnetic Reynolds number in the convection zone is large. Subsequently it was shown that a necessary condition for a velocity field to produce a fast dynamo action is to exhibit Lagrangian chaos or singularities \citep{Soward1994,Childress1995}, as can be expected for example in a turbulent flow. Extending the previous definitions, a dynamo is said to be furious ("very fast" in \citet{Alboussiere2020}) if the magnetic growth rate increases without upper bound with the magnetic Reynolds number, corresponding to magnetic growth on an even smaller time scale than advection. The three types of dynamo, slow, fast and furious are illustrated in Figure \ref{fig:illustration}.

Besides the anisotropic dynamo, among the simplest dynamos are the multicellular flow studied by \citet{Roberts1972} and the monocellular flow studied by \citet{Ponomarenko1973}, both being helical. In the Roberts dynamo, the fluid motion is smooth and not chaotic, leading to slow dynamo action. However, by adding singularities at the stagnation points of the flow it is possible to introduce an additional time scale which, if taken sufficiently small, can lead to fast dynamo action \citep{Soward1987}. In the Ponomarenko dynamo, depending whether the flow shear between the inner cylinder and the outer cylinder is smooth or infinite, the dynamo is either slow \citep{Ruzmaikin1988} or fast \citep{Gilbert1988}. 
Even in the fast case, magnetic diffusion appears to be a crucial ingredient of the Ponomarenko dynamo, as it is the only way to generate the radial component of the magnetic field from its azimuthal component \citep{Gilbert1988}. Similarly, in the anisotropic dynamo, magnetic diffusion is also crucial. However, as will be shown below, anisotropy now helps to generate both the radial and azimuthal components of the magnetic field, turning the dynamo into a fast or furious process depending on the type of shear that is considered.

\section{General formulation}
\label{sec:General formulation}

\begin{figure}
\begin{center}
\includegraphics[scale=0.4]{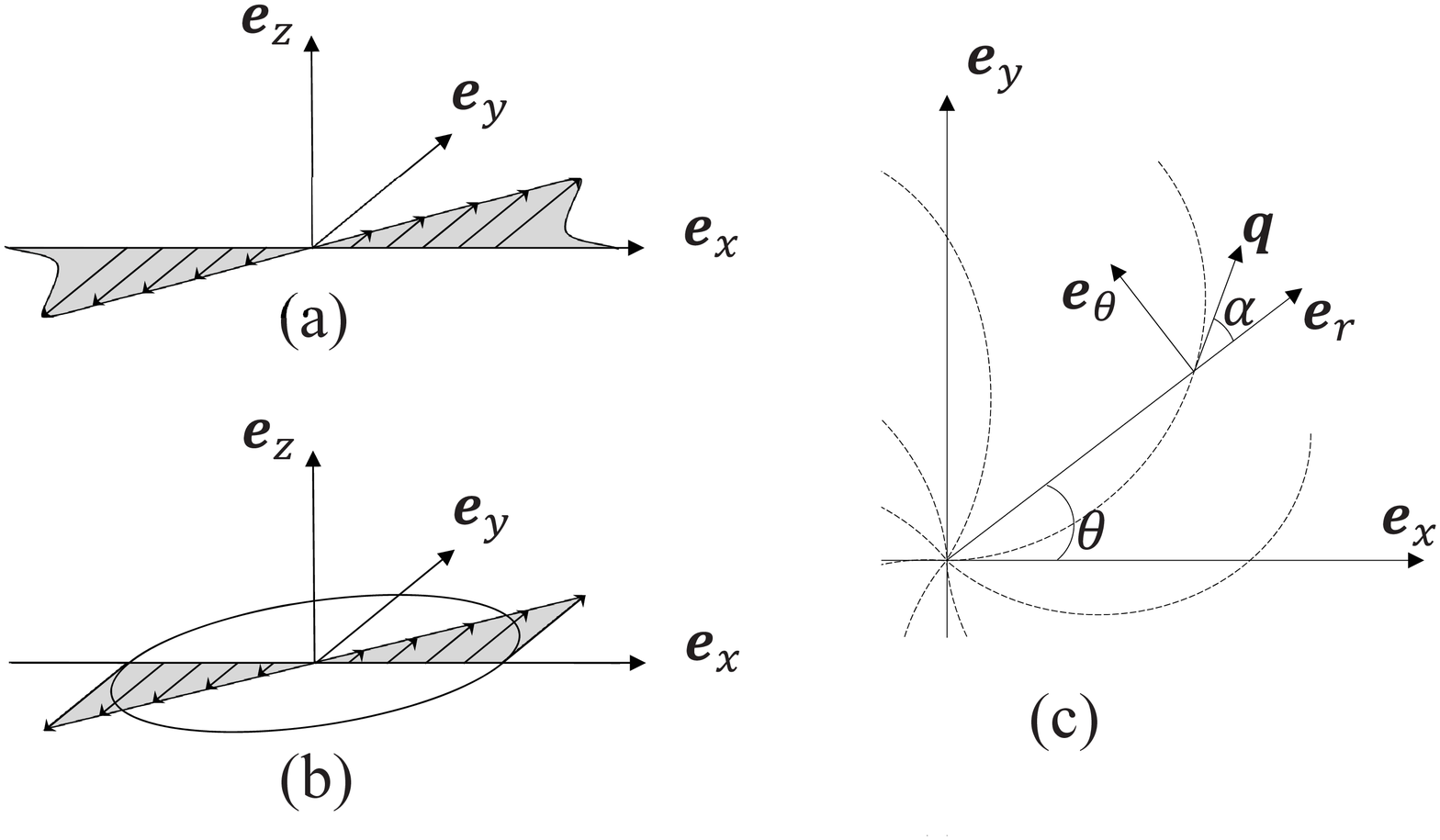}
\caption{Left: Illustration of the velocity field given by differential rotation in the horizonthal plane, with (a) a smooth shear and (b) an infinite shear. Right: Sketch of the logarithmic spirals tangential to the vector $\bq$.}
\label{fig:flow}
\end{center}
\end{figure}

We will consider two velocity fields, corresponding to differential rotation with either smooth or infinite shear as illustrated in Figures \ref{fig:flow}a and \ref{fig:flow}b. In cylindrical coordinates $(r,\theta,z)$, the smooth velocity field  is given by 
\begin{equation}
\bU = (0,r\Omega(r),0),
\label{eq:smooth shear}
\end{equation}
where the angular velocity $\Omega(r)$ is a continuous and differentiable function of $r$. 
The velocity field with infinite shear is given by
\begin{equation}
\bU=
\left\{
\begin{split}
r\Omega \be_{\theta}&,&\text{for}\; r< R \\
0&,&\text{for} \; r>R
\end{split}\;\;\;,
\right. \label{eq:infinite shear}
\end{equation}
where $(\be_r,\be_{\theta},\be_z)$ is the cylindrical coordinate system. The motion described by (\ref{eq:infinite shear}) corresponds to a solid body rotation of an inner-cylinder of radius $R$ with the angular velocity $\Omega$, surrounded by a medium at rest.

The assumption that electrical conductivity, or magnetic permeability, is anisotropic means that it takes a different value depending on the direction considered.
Following \citet{Ruderman1984}, the electrical conductivity and magnetic permeability are defined by $\sigma_{\parallel}$ and $\mu_{\parallel}$ in a given direction $\bq$, and by $\sigma_{\perp}$ and $\mu_{\perp}$ in the directions perpendicular to $\bq$, with $\bq$ a unit vector. 
In the direction parallel to $\bq$,  Ohm's law and the relation between $\bH$ and $\bB$  are written in the form $\bJ\cdot \bq=\sigma_{\parallel} (\bE\cdot \bq)$ and $\bB\cdot \bq=\mu_{\parallel}(\bH\cdot \bq)$, while
in the directions perpendicular to $\bq$,  they  are written as $\bJ-(\bJ\cdot\bq)\bq=\sigma_{\perp} (\bE-(\bE\cdot\bq)\bq)$ and $\bB-(\bB\cdot\bq)\bq=\mu_{\perp}(\bH-(\bH\cdot\bq)\bq)$.
This leads to two symmetric tensors, $\lbrack\sigma_{ij}\rbrack$ for the electrical conductivity and $\lbrack\mu_{ij}\rbrack$ for the magnetic permeability, defined by
\refstepcounter{equation}
$$
\lbrack\sigma_{ij}\rbrack=\sigma_{\perp} \delta_{ij} + (\sigma_{\parallel}-\sigma_{\perp})q_i q_j, \;\;\;\;\;\;\;\;\;
\lbrack\mu_{ij}\rbrack=\mu_{\perp} \delta_{ij} + (\mu_{\parallel}-\mu_{\perp})q_i q_j. 
  \eqno{(\theequation{\mathit{a,}\mathit{b}})}\label{eq:conductivity permeability tensors}
$$

In the magnetohydrodynamic approximation, Maxwell's equations
and Ohm's law take the following forms
\begin{subequations}\label{eq:MaxwellOhm}
\begin{align}
\bH&=\lbrack\mu_{ij}\rbrack^{-1}\bB,\label{eq:Maxwell1}\\
\nabla \cdot \bB &= 0, \label{eq:Maxwell2}\\
 \bJ&=\nabla \times \bH, \label{eq:Ampere}\\
 \partial_t \bB &= - \nabla \times \bE\label{eq:Faraday}\\
\bJ&=\lbrack\sigma_{ij}\rbrack(\bE + \bU \times \bB),\label{eq:Ohm}
\end{align}
\end{subequations}

leading to the equation for the magnetic induction $\bB$,
\begin{equation}
\partial_t \bB = \nabla \times (\bU \times \bB) - \nabla \times \left( \lbrack\sigma_{ij}\rbrack^{-1} \nabla \times \left( \lbrack\mu_{ij}\rbrack^{-1}\bB \right) \right),
\label{eq:induction equation}
\end{equation}
where
\refstepcounter{equation}
$$
\lbrack\sigma_{ij}\rbrack^{-1}=\sigma_{\perp}^{-1} \left(\delta_{ij} + \sigma q_i q_j \right), \;\;\;\;\;\;\;\;\;
\lbrack\mu_{ij}\rbrack^{-1}= \mu_{\perp}^{-1} \left(\delta_{ij} + \mu q_i q_j \right),
  \eqno{(\theequation{\mathit{a,}\mathit{b}})}\label{eq:resistivity and inv permeability tensors}
$$
with 
\refstepcounter{equation}
$$
\sigma=\frac{\sigma_{\perp}}{\sigma_{\parallel}}-1, \;\;\;\;\;\;\;\;\;
\mu= \frac{\mu_{\perp}}{\mu_{\parallel}}-1.
  \eqno{(\theequation{\mathit{a,}\mathit{b}})}\label{eq:sigma mu}
$$

As in \citet[2021]{Plunian2020}, we choose $\bq$ as a unit vector in the horizontal plane defined by
 \begin{equation}
 \bq=c \; \be_r+s \; \be_{\theta},
 \end{equation}
 where $c=\cos\alpha$, $s=\sin\alpha$, with $\alpha$ a prescribed angle. 
The vector $\bq$ is tangent to logarithmic spirals in the horizonthal plane ($\be_r$,$\be_{\theta}$), as illustrated in Figure \ref{fig:flow}c. 

Since the velocity is stationary and independent of $z$, and as we are considering only axisymmetric solutions, we can look for the magnetic induction in the form
\begin{equation}
\bB=\bB(r)\exp(\gamma t + \ii kz),
\label{eq:magnetic induction}
\end{equation}
where $\bB(r)$ is the axisymmetric magnetic mode at vertical wave number $k$. In (\ref{eq:magnetic induction}) a positive value of the real part of the magnetic growth rate $\gamma$ is the signature of dynamo action, the dynamo threshold corresponding to $\Re\{\gamma\}=0$.

Replacing (\ref{eq:smooth shear}) and (\ref{eq:magnetic induction}) in (\ref{eq:induction equation}), and after some algebra (see Appendix \ref{sec: eq of Br and Bteta}), one obtains the following equations for $B_r(r)$ and $B_{\theta}(r)$,
\begin{subequations}\label{eq:gammaBrBteta}
\begin{align}
\gamma B_r &= -\varepsilon \left[ \mu c^2 k^2 B_r + (1+\sigma s^2) D_k(B_r) - cs(\sigma -\mu) k^2 B_{\theta} \right], \label{eq:gammaBr}\\
\gamma B_{\theta}  &= -\varepsilon \left[ \sigma c^2 k^2 B_{\theta} + (1+\mu s^2)D_k (B_{\theta}) - cs(\sigma -\mu) D_k(B_r) \right] + r \Omega'(r) B_r \label{eq:gammaBteta},
\end{align}
\end{subequations}
where $\varepsilon = (\sigma_{\perp}\mu_{\perp})^{-1}$, and 
\begin{equation}
D_{k}(X)=k^2X-\partial_r\left(\frac{1}{r}\partial_r(rX)\right). 
\end{equation}
Normalizing the distance by some value $a$, and time by $|\Omega^{-1}(a)|$, corresponds in (\ref{eq:gammaBrBteta}$\mathit{a,}\mathit{b}$) to
replace $\varepsilon$ by the inverse of the magnetic Reynolds number 
\begin{equation}
\Rm=\sigma_{\perp}\mu_{\perp} a^2 |\Omega(a)|,
\label{eq:Rm}
\end{equation}
the fast dynamo problem refering to $\Rm \gg 1$, or equivalently to $\varepsilon \ll 1$. 

In Section \ref{sec:smooth differential rotation}, an asymptotic analysis of (\ref{eq:gammaBrBteta}$\mathit{a,}\mathit{b}$) for $\varepsilon \ll 1$, will 
allow to estimate the leading order of the magnetic growth rate in the case of a smooth shear given by (\ref{eq:smooth shear}). On the other hand, this cannot be done as easily for the case of a solid body rotation given by (\ref{eq:infinite shear}). Indeed, as $\Omega'(r)=0$ in both regions $r<R$ and $r>R$, the system (\ref{eq:gammaBrBteta}$\mathit{a,}\mathit{b}$) reduces to two anisotropic diffusion equations, without velocity term. Reminding that dynamo action is a conversion of kinetic into magnetic energy, the system (\ref{eq:gammaBrBteta}$\mathit{a,}\mathit{b}$) is therefore not sufficient to describe the dynamo process.  In fact, we will see that the velocity is only involved in the boundary conditions accross $r=R$. Therefore, it will be necessary to solve (\ref{eq:gammaBrBteta}$\mathit{a,}\mathit{b}$) with appropriate boundary conditions, in order to derive the magnetic growth rate $\gamma$ and study its behaviour for $\varepsilon \ll 1$. This will be the subject of Section \ref{sec:solid body}.

\section{Fast dynamo for smooth differential rotation}
\label{sec:smooth differential rotation}
Here we follow a similar line of arguments to the one developed for the smooth Ponomarenko dynamo \citep{Gilbert1988, Gilbert2003}, essentially based on a boundary analysis. 
In the asymptotic limit $\varepsilon \ll 1$ we expand $\gamma$, $B_r$ and $B_{\theta}$ in powers of $\varepsilon^{1/2}$, such that
\begin{eqnarray}
\gamma &=& \gamma_0 + \varepsilon^{\frac{1}{2}} \gamma_1 + \varepsilon \gamma_2+ \ldots, \label{eq:expgamma}\\
 B_r &=& B_{r0} + \varepsilon^{\frac{1}{2}} B_{r1}+ \varepsilon B_{r2} + \ldots, \\
 B_{\theta} &=& B_{\theta0} + \varepsilon^{\frac{1}{2}} B_{\theta1} + \varepsilon B_{\theta2} + \ldots,\label{eq:expBteta}
\end{eqnarray}
and we set
\begin{equation}
k=K \varepsilon^{-\frac{1}{2}}\;\;\; \text{and}\;\;\; r=a+\varepsilon^{\frac{1}{2}} \zeta,
\end{equation}
meaning that we search for a magnetic mode at some radius $r=a$ within a magnetic boundary layer.
The $r$-derivative takes the form $\partial_r= \varepsilon^{-\frac{1}{2}}\partial_\zeta$, leading to
\begin{equation}
\frac{1}{r}\partial_r(rX)=\frac{X}{a} + \varepsilon^{-\frac{1}{2}}\frac{\partial X}{\partial \zeta}\;\;\; \text{and}\;\;\; D_{k}(X)=\varepsilon^{-1}(K^2-\frac{\partial^2}{\partial \zeta^2})X  -\varepsilon^{-\frac{1}{2}}\frac{1}{a}\frac{\partial X}{\partial \zeta}+\frac{X}{a^2},
\label{eq:exppartial}
\end{equation}
where $X$ can be any variable, $e.g.$ $B_r$ or $B_{\theta}$.
Rewriting (\ref{eq:Maxwell2})  as
\begin{equation}
B_z = \ii k^{-1} \frac{1}{r}\partial_r(rB_r),
\label{eq:divB2}
\end{equation}
and using (\ref{eq:exppartial}a), we find
\begin{equation}
B_z = B_{z0} + \varepsilon^{\frac{1}{2}} B_{z1}+ \varepsilon B_{z2} + \ldots,
\label{eq:expBz}
\end{equation}
where, at leading order, $B_z=B_{z0}=\ii K^{-1}\partial B_{r0} / \partial \zeta$.
A striking difference with the Ponomarenko dynamo is that, at leading order, none of the three components $B_{r0}, B_{\theta0}$ and $B_{z0}$ is identically zero, whereas in the Ponomarenko dynamo $B_{r0}=0$. 

Assuming that the variations in $r$ are of the same order of magnitude as those in $z$, 
we can approximate 
$D_k(X)\approx \varepsilon^{-1}K^2 X_0$.
Replacing (\ref{eq:expgamma}-\ref{eq:expBteta}) in (\ref{eq:gammaBrBteta}$\mathit{a,}\mathit{b}$) then leads, at leading order, to the equations
\begin{subequations}\label{eq:gammaBrBteta0}
\begin{align}
\left[\gamma_0 + (1+\sigma s^2 + \mu c^2) K^2\right] B_{r0} - cs(\sigma -\mu) K^2 B_{\theta0} &= 0, \label{eq:gammaBr0}\\
\left[ cs(\sigma -\mu) K^2 + a \Omega'(a) \right] B_{r0} - \left[\gamma_0 + (1+\sigma c^2 + \mu s^2) K^2 \right] B_{\theta0} &= 0 \label{eq:gammaBteta0}.
\end{align}
\end{subequations}
In (\ref{eq:gammaBrBteta0}$\mathit{a,}\mathit{b}$), looking for non-zero $B_{r0}$ and $B_{\theta0}$ leads to the following expression for the leading order growth rate,
\begin{equation}
\gamma_0 = \frac{K^2}{2}\left[ -(\sigma + \mu + 2) \pm |\sigma - \mu|\left( 1 + \frac{4csa\Omega'(a)}{K^2(\sigma - \mu)}\right)^{1/2} \right].
\label{eq:gamma0}
\end{equation}
A necessary condition for dynamo action is $\gamma_0 > 0$, which corresponds to
\begin{equation}
a \Omega'(a) > \frac{K^2(\sigma +1)(\mu+1)}{cs(\sigma - \mu)}\equiv \frac{K^2}{cs}\left( \frac{\mu_{\parallel}}{\mu_{\perp}}-\frac{\sigma_{\parallel}}{\sigma_{\perp}}\right)^{-1}.
\label{eq:threshold smooth}
\end{equation}
In (\ref{eq:threshold smooth}), we note that, from (\ref{eq:sigma mu}$\mathit{a,}\mathit{b}$), we have $\sigma+1>0$ and $\mu+1>0$.
Then, assuming $cs(\sigma-\mu)>0$, (\ref{eq:threshold smooth}) implies that the derivative of $\Omega(r)$ at $r=a$ must be positive and sufficiently large. This can be achieved in different ways, one of them being $\Omega(r) <0$ and $\lim\limits_{r \rightarrow \infty}\Omega(r)=0$. 
Although here the differential rotation is smooth, this picture is consistent with the one obtained for an infinite shear \citep{Plunian2021}.
We note that in (\ref{eq:gamma0}), swapping $\sigma$ and $\mu$, and changing $\Omega(r)$ in $-\Omega(r)$, does not change the result, extending the duality argument put forward by \cite{Favier2013} and \cite{Marcotte2021} to the cases of anisotropic electrical conductivity and anisotropic magnetic permeability.  

From (\ref{eq:gamma0}) and (\ref{eq:threshold smooth}) we conclude that, in the limit $\Rm\gg1$, the magnetic growth rate at leading order can be positive and independent of $\Rm$, making the smooth anisotropic dynamo a fast dynamo. This is true only if $\sigma \ne \mu$, meaning that the degree of anisotropy of electrical conductivity must be different from the one of magnetic permeability.  An additional condition is that $cs\ne 0$, which means that the two limiting cases of geometry anisotropy, namely straight radii and circles, must be excluded.

\section{Furious dynamo for infinite shear}
\label{sec:solid body}
\subsection{Renormalization and boundary conditions}
\label{sec:renormalization and BC}
In the case of an inner cylinder in solid body rotation surrounded by a medium at rest, given by the velocity (\ref{eq:infinite shear}), the system to solve is identical in each region $r<R$ and $r>R$, given by (\ref{eq:gammaBrBteta}$\mathit{a,}\mathit{b}$) with $\Omega'(r)=0$.
Again, normalizing the distance and time by respectively $R$ and $|\Omega|^{-1}$,
 leads to
\begin{subequations}\label{eq:gammaBrBteta2}
\begin{align}
\tilde{\gamma} B_r &= - \left[ \mu c^2k^2  B_r + (1+\sigma s^2)  D_k(B_r) - cs(\sigma -\mu)k^2 B_{\theta} \right] \label{eq:gammaBr2}\\
\tilde{\gamma} B_{\theta}  &= - \left[ \sigma c^2k^2 B_{\theta} + (1+\mu s^2)D_k (B_{\theta}) - cs(\sigma -\mu) D_k(B_r) \right] \label{eq:gammaBteta2},
\end{align}
\end{subequations}
where 
\begin{equation}
\tilde{\gamma} = \gamma \Rm,
\label{eq:gammatildeRm}
\end{equation}
with $\Rm$ defined by (\ref{eq:Rm}), replacing $a$ by $R$, and $\Omega(a)$ by $\Omega$.
The system of equations (\ref{eq:gammaBrBteta2}$\mathit{a,}\mathit{b}$) must be completed by the appropriate boundary conditions for $r=0$ and $r\rightarrow \infty$,
\refstepcounter{equation}
$$
  B_r(r=0)= B_{\theta}(r=0)=
\lim_{r\rightarrow \infty}B_r=\lim_{r\rightarrow \infty}B_{\theta} = 0,
  \eqno{(\theequation{\mathit{a-}\mathit{d}})}\label{eq:r0infinite}
$$
and by the continuity across $r=1$ of the normal component of $\bB$, and of the tangential components of $\bH$ and $\bE$,
\refstepcounter{equation}
$$
  \left[B_r\right]^{r=1^+}_{r=1^-} = \left[H_{\theta}\right]^{r=1^+}_{r=1^-} =\left[H_z\right]^{r=1^+}_{r=1^-} =\left[E_{\theta}\right]^{r=1^+}_{r=1^-} =\left[E_z\right]^{r=1^+}_{r=1^-} =0,
  \eqno{(\theequation{\mathit{a-}\mathit{e}})}\label{eq:r1}
$$
where $\left[X\right]^{r=1^+}_{r=1^-} =X(r=1^+)-X(r=1^-)$.
From (\ref{eq:Maxwell1}) and (\ref{eq:divB2}), (\ref{eq:r1}$\mathit{a-}\mathit{c}$) can be rewritten as
\refstepcounter{equation}
$$
\left[B_r\right]^{r=1^+}_{r=1^-} = \left[B_{\theta}\right]^{r=1^+}_{r=1^-} =\left[\partial_r B_r\right]^{r=1^+}_{r=1^-} =0,
\eqno{(\theequation{\mathit{a-}\mathit{c}})}
\label{eq:r12}
$$
meaning that $B_r$, $B_{\theta}$ and the derivative of $B_r$  are continuous across $r=1$.
From (\ref{eq:Faraday}) we have $E_{\theta}=-\ii k^{-1}\tilde{\gamma}B_r$, implying that
the two conditions (\ref{eq:r1}$\mathit{a}$) and (\ref{eq:r1}$\mathit{d}$) are redundant. As for the last one (\ref{eq:r1}$\mathit{e}$), using (\ref{eq:Ohm}) it can be rewritten  as
\begin{equation}
\left[J_z\right]^{r=1^+}_{r=1^-} = \overline{\Rm} B_r(r=1),
\label{eq:Ezcont1}
\end{equation}
where $\bJ$ has been normalized by $(\mu_{\perp}R)^{-1}$, and $\overline{\Rm} $ is still the magnetic Reynolds number except that it is signed, keeping track of the direction of the rotation, anticlockwise ($\overline{\Rm}>0$) or clockwise ($\overline{\Rm}<0$). It is defined by $\overline{\Rm} = \sign(\Omega) \cdot \Rm $. 

\subsection{Resolution}
\label{sec:resolution}
The resolution of the system (\ref{eq:gammaBrBteta2}$\mathit{a,}\mathit{b}$) follows the same line of reasoning as that of \citet{Plunian2021} except that here, instead of the dynamo threshold corresponding to $\tilde{\gamma}=0$,  we solve the system for any value of $\tilde{\gamma}$.

Introducing
\refstepcounter{equation}
$$
k_{\sigma}=k\left(\frac{1+\sigma+\tilde{\gamma}/k^2}{1+\sigma s^2}\right)^{1/2}, \quad k_{\mu}=k\left(\frac{1+\mu+\tilde{\gamma}/k^2}{1+\mu s^2}\right)^{1/2},
\eqno{(\theequation{\mathit{a},\mathit{b}})}\label{eq:ksigma kmu}
$$
and with the help of the identity
\begin{equation}
D_{k_1}(X)=D_{k_2}(X) +(k_1^2-k_2^2)X, \label{eq:identity}
\end{equation}
the system (\ref{eq:gammaBrBteta2}$\mathit{a,}\mathit{b}$) takes the following form
\begin{subequations}\label{eq:gammaBrBteta3}
\begin{align}
(1+\sigma s^2)  D_{k_{\sigma}}(B_r) &= (\sigma -\mu)ck^2 (cB_r + s B_{\theta}) \label{eq:gammaBr3}\\
(1+\mu s^2)D_{k_{\mu}} (B_{\theta})  &=(\sigma -\mu)c \left( s D_k(B_r)-ck^2 B_{\theta} \right) \label{eq:gammaBteta3}.
\end{align}
\end{subequations}
Then we can show that (see Appendix \ref{sec:derivation of DoD})
\refstepcounter{equation}
$$
D_{k_{\mu}}(cB_r + s B_{\theta})=D_{k_{\sigma}}\left( s D_k(B_r)-ck^2 B_{\theta} \right)=0.
\eqno{(\theequation{\mathit{a},\mathit{b}})}\label{eq:Dkk}
$$
Then, using (\ref{eq:Dkk}$\mathit{a,}\mathit{b}$) and  (\ref{eq:gammaBrBteta3}$\mathit{a,}\mathit{b}$)
leads to
\refstepcounter{equation}
$$
\left(D_{k_{\mu}}\circ D_{k_{\sigma}}\right)(B_r)=\left(D_{k_{\sigma}}\circ D_{k_{\mu}}\right)(B_{\theta}) = 0.
\eqno{(\theequation{\mathit{a},\mathit{b}})}\label{eq:differential equations}
$$
The two operators $D_{k_{\sigma}}$ and $D_{k_{\mu}}$ being commutative, $B_r$ and $B_{\theta}$ satisfy the same 
linear differential equation of fourth order.  
As the solution of $D_{\nu}(X)=0$ is a linear combination of $I_1(\nu r)$ and $K_1(\nu r)$, where $I_1$ and $K_1$ are first and second kind modified Bessel functions of order 1, the solutions of (\ref{eq:differential equations}$\mathit{a,}\mathit{b}$)
are a linear combination of $I_1(k_{\sigma} r)$, $K_1(k_{\sigma} r)$, $I_1(k_{\mu} r)$ and $K_1(k_{\mu} r)$.
Applying the boundary conditions (\ref{eq:r0infinite}$\mathit{a-}\mathit{d}$) and (\ref{eq:r12}$\mathit{a,}\mathit{b}$), $B_r$ and $B_{\theta}$ can be written in the following form,
\begin{eqnarray}
B_r=&&
\left\{
\begin{split}
r< 1,&\;\;\;\;\;-s\left(\lambda_{\sigma} \frac{I_1(k_{\sigma} r)}{I_1(k_{\sigma})}+ \lambda_{\mu} \frac{I_1(k_{\mu} r)}{I_1(k_{\mu})}\right)&  \\
r> 1,&\;\;\;\;\;-s\left(\lambda_{\sigma} \frac{K_1(k_{\sigma} r)}{K_1(k_{\sigma})}+ \lambda_{\mu} \frac{K_1(k_{\mu} r)}{K_1(k_{\mu})}\right)&
\end{split}
\right.  \label{eq:Brlambda}
\end{eqnarray}

\begin{eqnarray}
B_{\theta}=&&
\left\{
\begin{split}
r< 1,&\;\;\;c \left(\lambda_{\sigma} \frac{I_1(k_{\sigma} r)}{I_1(k_{\sigma})}+ \frac{\mu s^2+(\tilde{\gamma} s^2)/(c^2k^2)}{1+\mu s^2}\lambda_{\mu} \frac{I_1(k_{\mu} r)}{I_1(k_{\mu})}\right) &  \\
r> 1,&\;\;\;c \left(\lambda_{\sigma} \frac{K_1(k_{\sigma} r)}{K_1(k_{\sigma})}+ \frac{\mu s^2+(\tilde{\gamma} s^2)/(c^2k^2)}{1+\mu s^2}\lambda_{\mu} \frac{K_1(k_{\mu} r)}{K_1(k_{\mu})} \right)&
\end{split}
\right. , \label{eq:Btetalambda}
\end{eqnarray}
where $B_{\theta}$ has been obtained from $B_r$ by replacing (\ref{eq:Brlambda}) in (\ref{eq:gammaBr3}).  
To do this, we need to calculate $D_{k_{\sigma}}(B_r)$, which is derived in Appendix \ref{sec:Derivation de Bteta}.

The continuity of  $\partial_rB_r$ (calculated in Appendix \ref{sec:derivation of BC}) across $r=1$, given by (\ref{eq:r12}$\mathit{c}$), leads to the additional identity between $\lambda_{\sigma}$ and $\lambda_{\mu}$
 \begin{equation}
 \lambda_{\sigma} \Gamma(k_{\sigma}) + \lambda_\mu \Gamma(k_{\mu})=0,
 \label{eq:A'cont}
 \end{equation}
 with
 \begin{equation}
 \Gamma(x)=x\left(\frac{I_0(x)}{I_1(x)} + \frac{K_0(x)}{K_1(x)}\right) \equiv \left(I_1(x) K_1(x)\right)^{-1},
 \label{eq:Gamma}
 \end{equation}
the last identity being the Wronskian relation 
\begin{equation}
I_m(x)K_{m+1}(x)+I_{m+1}(x)K_m(x)=1/x. \label{eq:Wronskian}
\end{equation}

Finally, to apply the last boundary condition (\ref{eq:Ezcont1}) we need to calculate the $z$-component of the current density, that is derived by replacing $B_r$ and $B_{\theta}$ given by (\ref{eq:Brlambda}) and (\ref{eq:Btetalambda}), in (\ref{eq:Maxwell1}) and (\ref{eq:Ampere}), leading to (see Appendix \ref{sec:derivation of j})
\begin{eqnarray}
\bJ_z=&&
\left\{
\begin{split}
r< 1,&\;\;\;c \left[ k_{\sigma} \lambda_{\sigma}\frac{I_0(k_{\sigma} r)}{I_1(k_{\sigma})} + \frac{s^2\tilde{\gamma}}{c^2k^2}\lambda_\mu k_\mu \frac{I_0(k_{\mu} r)}{I_1(k_{\mu})}\right]  \\
r> 1,&\;\;\;-c \left[ k_{\sigma} \lambda_{\sigma}\frac{K_0(k_{\sigma} r)}{K_1(k_{\sigma})} + \frac{s^2\tilde{\gamma}}{c^2k^2}\lambda_\mu k_\mu \frac{K_0(k_{\mu} r)}{K_1(k_{\mu})}\right]
\end{split}
\right. . \label{eq:jz}
\end{eqnarray}

Replacing (\ref{eq:Brlambda}) and (\ref{eq:jz}) in (\ref{eq:Ezcont1}), and using (\ref{eq:A'cont}), leads to the following dispersion relation
\begin{equation}
\overline{\Rm} = \frac{c}{s}\left(1-\frac{s^2\tilde{\gamma}}{c^2k^2}\right)\left(I_1(k_{\sigma}) K_1(k_{\sigma})-I_1(k_{\mu})K_1(k_{\mu})\right)^{-1}.
\label{eq:Omegac}
 \end{equation}
The dynamo threshold obtained for $\tilde{\gamma}=0$ has been the subject of a previous paper \citep{Plunian2021}.

\subsection{Asymptotic behaviour of $\tilde{\gamma}$ in the limit of large magnetic Reynolds numbers}
\label{sec:asymptotics}
We note that in (\ref{eq:Omegac}), as $k_{\sigma}$ and $k_{\mu}$ given by (\ref{eq:ksigma kmu}$\mathit{a,}\mathit{b}$) also depend on $\tilde{\gamma}$,
we cannot derive an explicit expression for $\tilde{\gamma}$.
Therefore to determine the asymptotic behaviour of $\tilde{\gamma}$ for $\Rm \gg 1$, two approaches are possible, either by solving numerically (\ref{eq:Omegac}) that we postpone to Subsection \ref{sec:Numerical solution}, or to carry out an asymptotic study, assuming that $k_{\sigma} \gg 1$ and $k_{\mu} \gg 1$. In the latter case, since \citep{Abramowitz1968}
\begin{equation}
\text{for} \quad |z|\gg1, \quad I_1(z)K_1(z) = \frac{1}{2z} + O(\frac{1}{z^3}),
\end{equation}
the dispersion relation (\ref{eq:Omegac}) can be written as
\begin{equation}
\overline{\Rm} \approx \frac{2c}{s}\left(1-\frac{s^2\tilde{\gamma}}{c^2k^2}\right) \frac{k_{\sigma}k_{\mu}}{k_{\mu}-k_{\sigma}}. \label{eq:Rmasymptotic}
\end{equation}
In (\ref{eq:Rmasymptotic}), replacing $k_{\sigma}$ and $k_{\mu}$ by their expressions (\ref{eq:ksigma kmu}$\mathit{a,}\mathit{b}$) leads to the following expression
\begin{equation}
\overline{\Rm} \approx \frac{-2k}{cs(\sigma-\mu)} g(\frac{\tilde{\gamma}}{k^2}), \label{eq:Rmg}
\end{equation}
where $g(x)$ is the function defined by
\begin{equation}
g(x)=(1+\sigma s^2)^{1/2}(1+\sigma+x)^{1/2}(1+\mu+x)+(1+\mu s^2)^{1/2}(1+\mu+x)^{1/2}(1+\sigma+x). \label{eq:g}
\end{equation}
For a given value of $\Rm$, as $\tilde{\gamma}$ depends on $k$, the maximum growth rate is obtained for $\partial \tilde{\gamma}/\partial k = 0$. Therefore, differentiating (\ref{eq:Rmg})  versus $k$, we find that this maximum is characterised by
\begin{equation}
g(x_0)=2x_0 g'(x_0), \label{eq:gpartial}
\end{equation}
where $g'$ is the derivative of $g$, and $x_0$ the solution of the following equation
\begin{eqnarray}
&&\left(\frac{1+\sigma s^2}{1+\sigma+x_0}\right)^{1/2}\left[2x_0^2+(1+\sigma)x_0-(1+\sigma)(1+\mu)\right] \nonumber\\
&+&\left(\frac{1+\mu s^2}{1+\mu+x_0}\right)^{1/2}\left[2x_0^2+(1+\mu)x_0-(1+\mu)(1+\sigma)\right]=0.
\label{eq:x}
\end{eqnarray}
\begin{figure}
\includegraphics[scale=0.25]{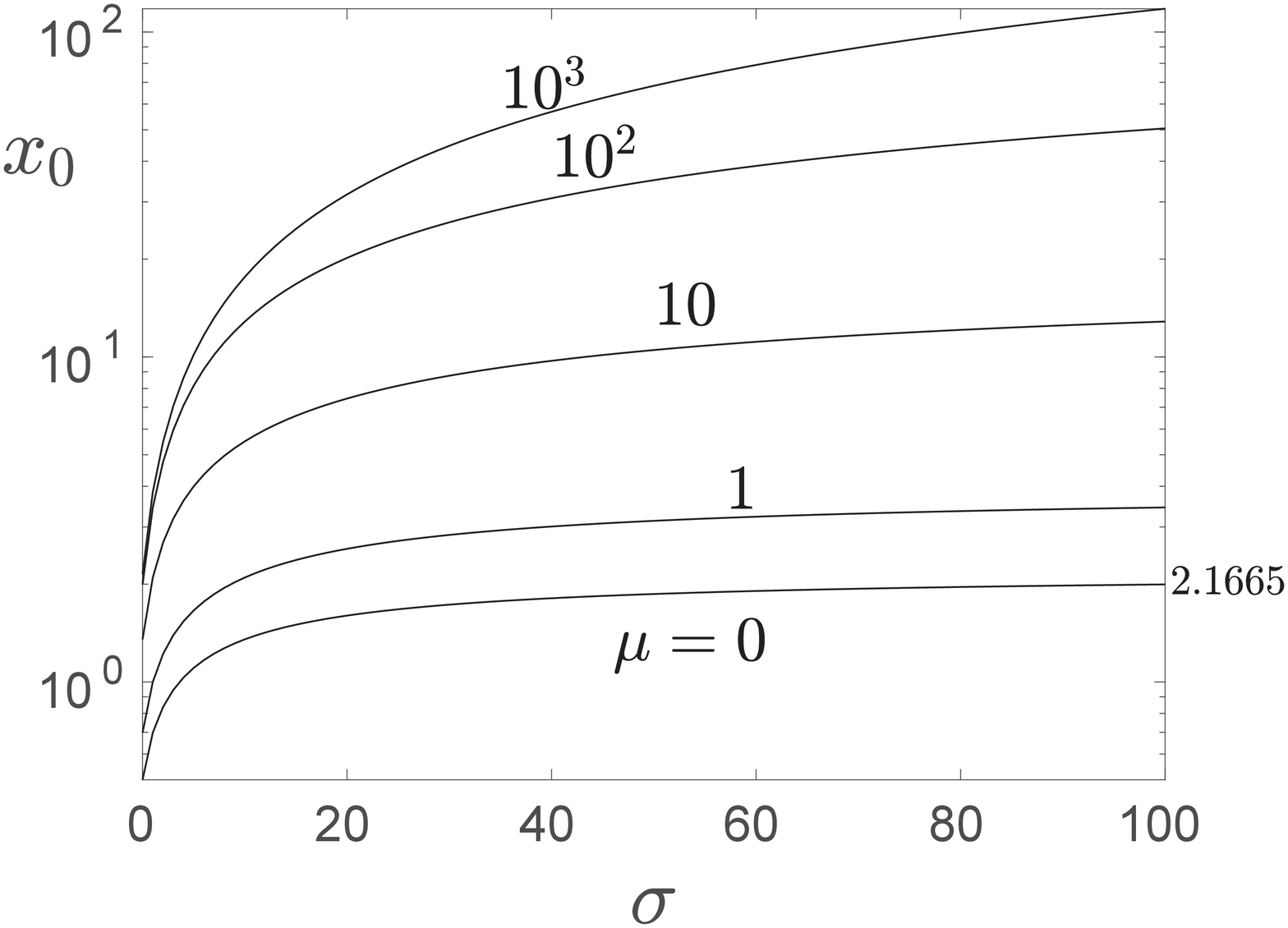}
\includegraphics[scale=0.25]{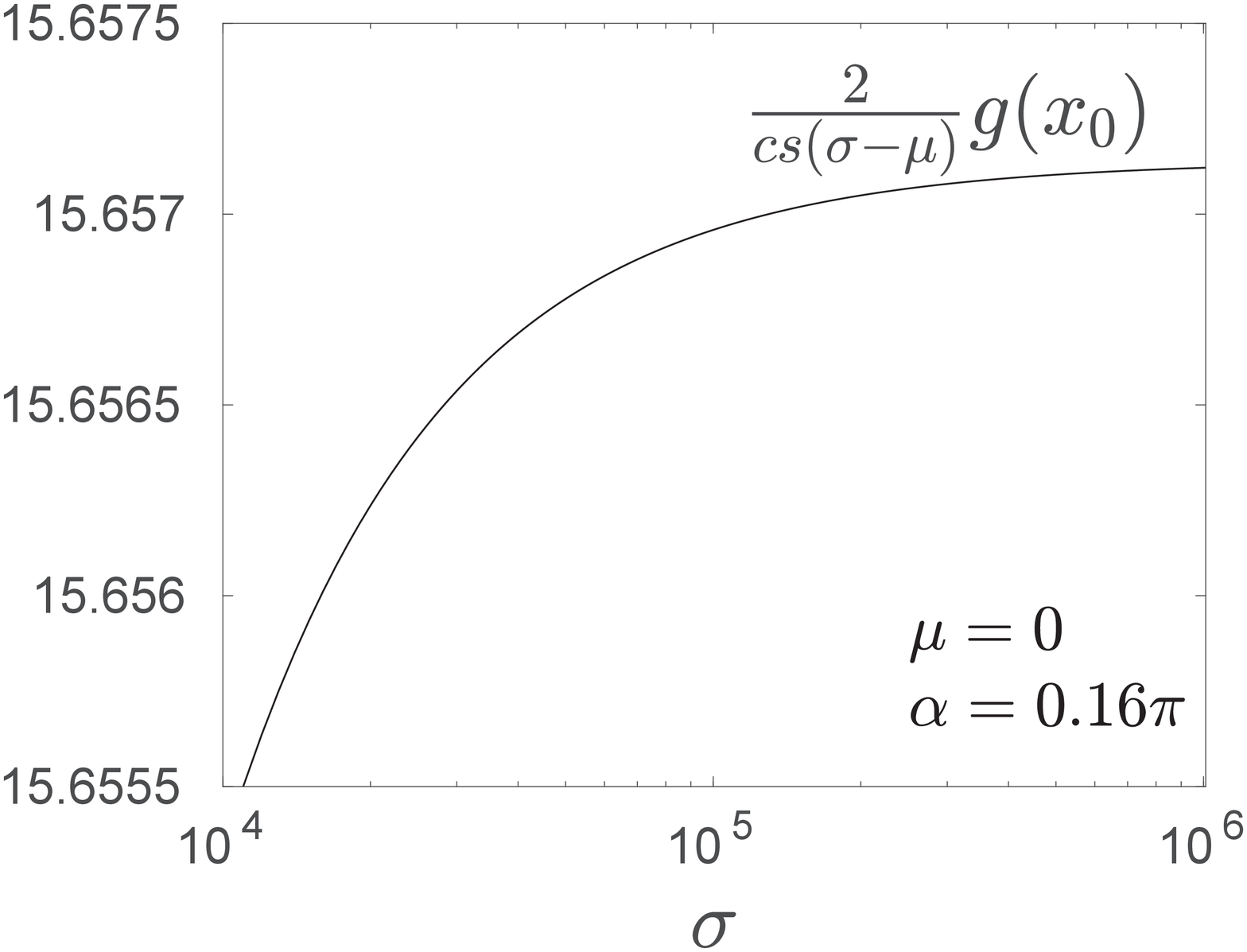}
\caption{(a) Plot of $x_0$ versus $\sigma$ for several values of $\mu$ and $\alpha=0.16 \pi$. (b) Plot of $\frac{2}{cs(\sigma-\mu)} g(x_0)$ versus $\sigma$ for $\mu=0$ and $\alpha=0.16 \pi$.}
\label{fig:x0}
\end{figure}
In Figure \ref{fig:x0}a, the solution  $x_0$ of (\ref{eq:x}) is plotted  versus $\sigma$, for $\alpha=0.16 \pi$ and several values of $\mu$. The value $\alpha=0.16 \pi$ is chosen in reference to the dynamo threshold minimum obtained for $\sigma \gg 1$ when $\mu=0$ \citep{Plunian2020}. For $\mu=0$ and $\sigma=10^6$ we find that $x_0 = 2.1665$, that will be used for comparison with the numerical results of Section \ref{sec:Numerical solution}. 
Replacing $x_0$ in (\ref{eq:g}), $\frac{2}{cs(\sigma-\mu)} g(x_0)$ is calculated and plotted in Figure \ref{fig:x0}b versus $\sigma$ for $\alpha=0.16 \pi$ and $\mu=0$. It takes the value 15.65715 for $\sigma=10^6$ which, again, will be used for comparison with the numerical results of Section \ref{sec:Numerical solution}.

As $x_0$ is entirely determined by $\sigma, \mu$ and $s$,
we conclude from (\ref{eq:Rmg}) that the maximum growth rate corresponds not only to $\tilde{\gamma}=x_0 k^2$ but also to $k\propto \Rm$, implying that $\tilde{\gamma}\propto \Rm^2$. From (\ref{eq:gammatildeRm}), we then conclude that
 \begin{equation}
\text{for}\; \Rm \gg 1, \quad \gamma\propto \Rm, \label{eq:gamma k Rm}
\end{equation}
making the anisotropic dynamo with infinite shear a furious dynamo.

Assuming $cs(\sigma-\mu) >0$, we note that (\ref{eq:Rmg}) implies that $k$ and $\overline{\Rm}$ have opposite signs, meaning that the dynamo action corresponds to a clockwise rotation of the inner cylinder. 
For $\Rm \gg 1$, as $k \propto \Rm$ and $\tilde{\gamma}/k^2 \rightarrow x_0$, from the definition of $k_{\sigma}$ and $k_{\mu}$ given in (\ref{eq:ksigma kmu}$\mathit{a,}\mathit{b}$), the two assumptions $k_{\sigma} \gg 1$ and $k_{\mu} \gg 1$ made at the beginning of Section \ref{sec:asymptotics}, are satisfied.

\subsection{Numerical solution of the dispersion relation (\ref{eq:Omegac})}
\label{sec:Numerical solution}
\begin{figure}
\begin{center}
\includegraphics[scale=0.4]{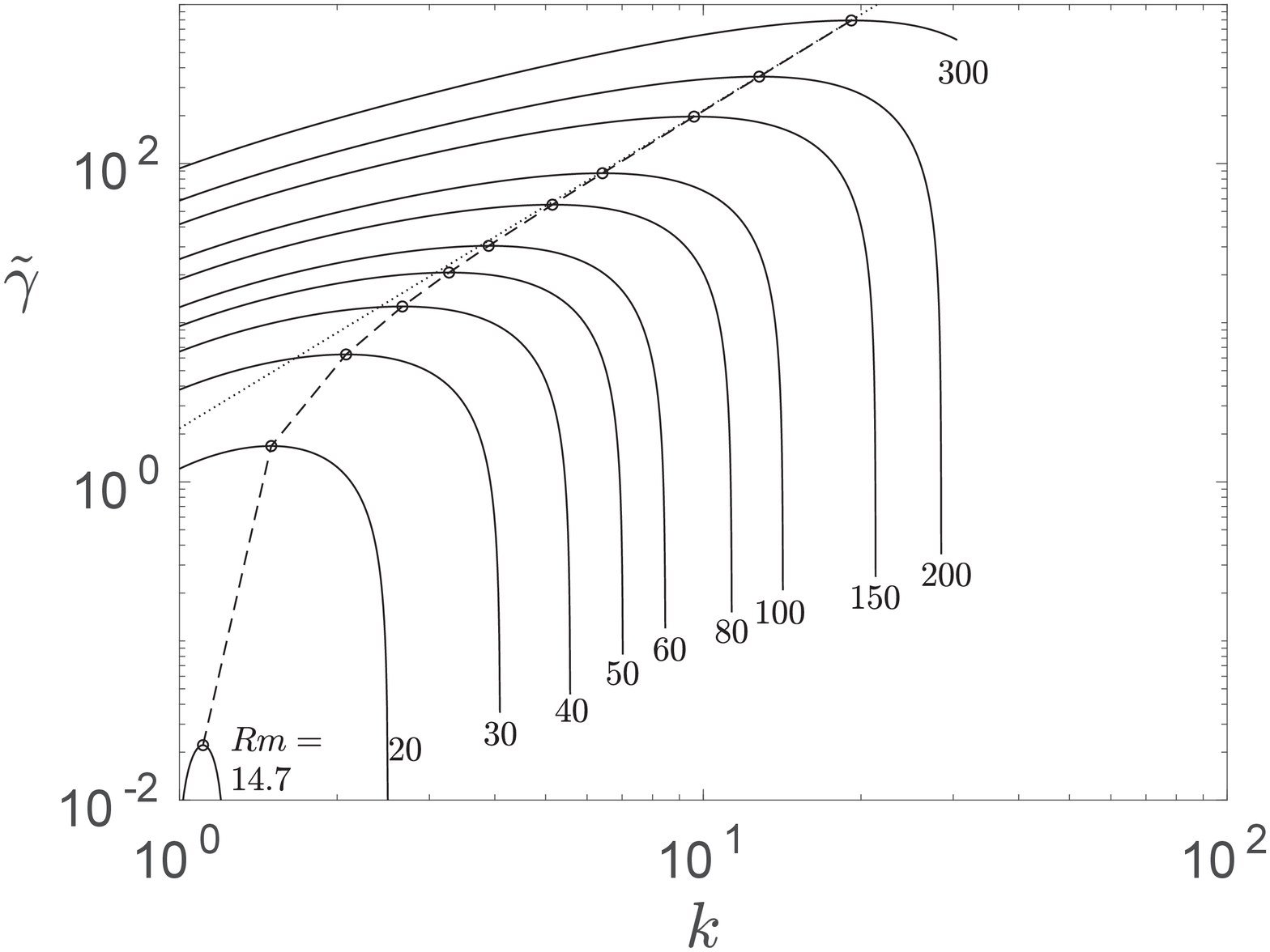}
\caption{Plot of $\tilde{\gamma}$ versus $k$ for $\mu=0$, $\sigma=10^6$, $\alpha=0.16 \pi$ and several values of $\Rm$. For each curve, the maximum $\tilde{\gamma}_0$ of $\tilde{\gamma}$, obtained for $k=k_0$, is denoted by a circle.}
\label{fig:gamma}
\end{center}
\end{figure}
In Figure \ref{fig:gamma}, the growthrate $\tilde{\gamma}$ obtained from (\ref{eq:Omegac}) is plotted versus $k$, for $\mu=0$, $\sigma=10^6$, $\alpha=0.16 \pi$ 
and several values of $\Rm$. As mentioned above, $\sigma$ is taken as sufficiently large in order to reach asymptotically the minimum dynamo threshold which, for $\mu=0$, is equal to $\Rm=14.7$ \citep{Plunian2020}. For other values of $\mu,\sigma$ or $\alpha$, the curves will be different, without however changing the asymptotic behaviour of $\tilde{\gamma}$ as  $\Rm \gg 1$. As found in Subsection \ref{sec:asymptotics}, the values of $\overline{\Rm}$ are found negative.

 In Figure \ref{fig:gamma}, for each curve, the maximum of $\tilde{\gamma}$ is denoted by a circle.
The dotted straight curve corresponds to $\tilde{\gamma}=x_0 k^2$ with $x_0=2.1665$ calculated from (\ref{eq:x}),
showing an excellent agreement between the asymptotic approach at large $\Rm$ and the numerical results.

\begin{figure}
\begin{center}
\includegraphics[scale=0.35]{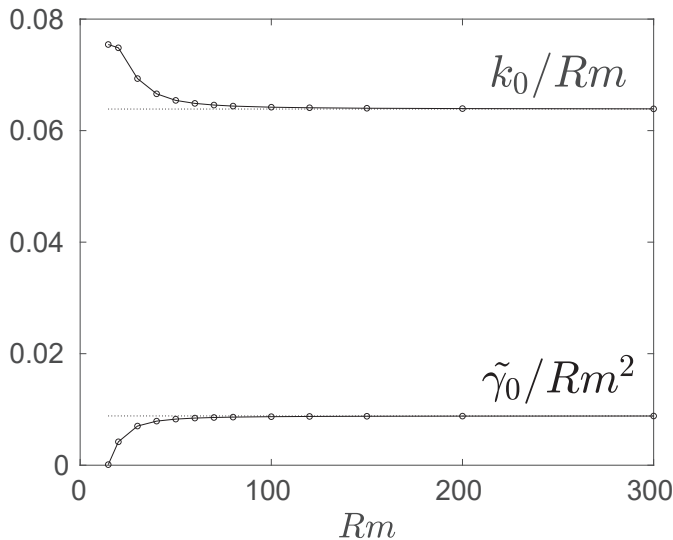}
\caption{(a) Plots of $k_0/\Rm$ and $\tilde{\gamma}_0/\Rm^2$ versus $\Rm$ for $\mu=0$, $\sigma=10^6$, $\alpha=0.16 \pi$. The horizonthal dotted lines correspond to the asymptotic approach of Section \ref{sec:asymptotics}.}
\label{fig:asymptotics}
\end{center}
\end{figure}

From Figure \ref{fig:gamma}, each maximum is denoted by its coordinates $(k_0,\tilde{\gamma}_0)$, such that
$\tilde{\gamma}_0= \max\limits_{k}\tilde{\gamma}(k)=\tilde{\gamma}(k_0)$. In Figure \ref{fig:asymptotics}a,
$k_0/\Rm$ and $\tilde{\gamma}_0/\Rm^2$  are then plotted versus $\Rm$. In the asymptotic limit $\Rm\gg1$
the scaling laws $k_0 \propto \Rm$ and $\tilde{\gamma}_0 \propto \Rm^2$ found in Section \ref{sec:asymptotics} are clearly validated. 
In addition, the horizonthal dotted lines corresponding to $\left(\frac{2}{cs(\sigma-\mu)}g(x_0)\right)^{-1}=1/15.65715$ and to $x_0\left(\frac{2}{cs(\sigma-\mu)}g(x_0)\right)^{-2}=2.1665/15.65715^2$, confirm the excellent agreement with the asymptotic approach. 

\begin{figure}
\begin{center}
\includegraphics[scale=0.22]{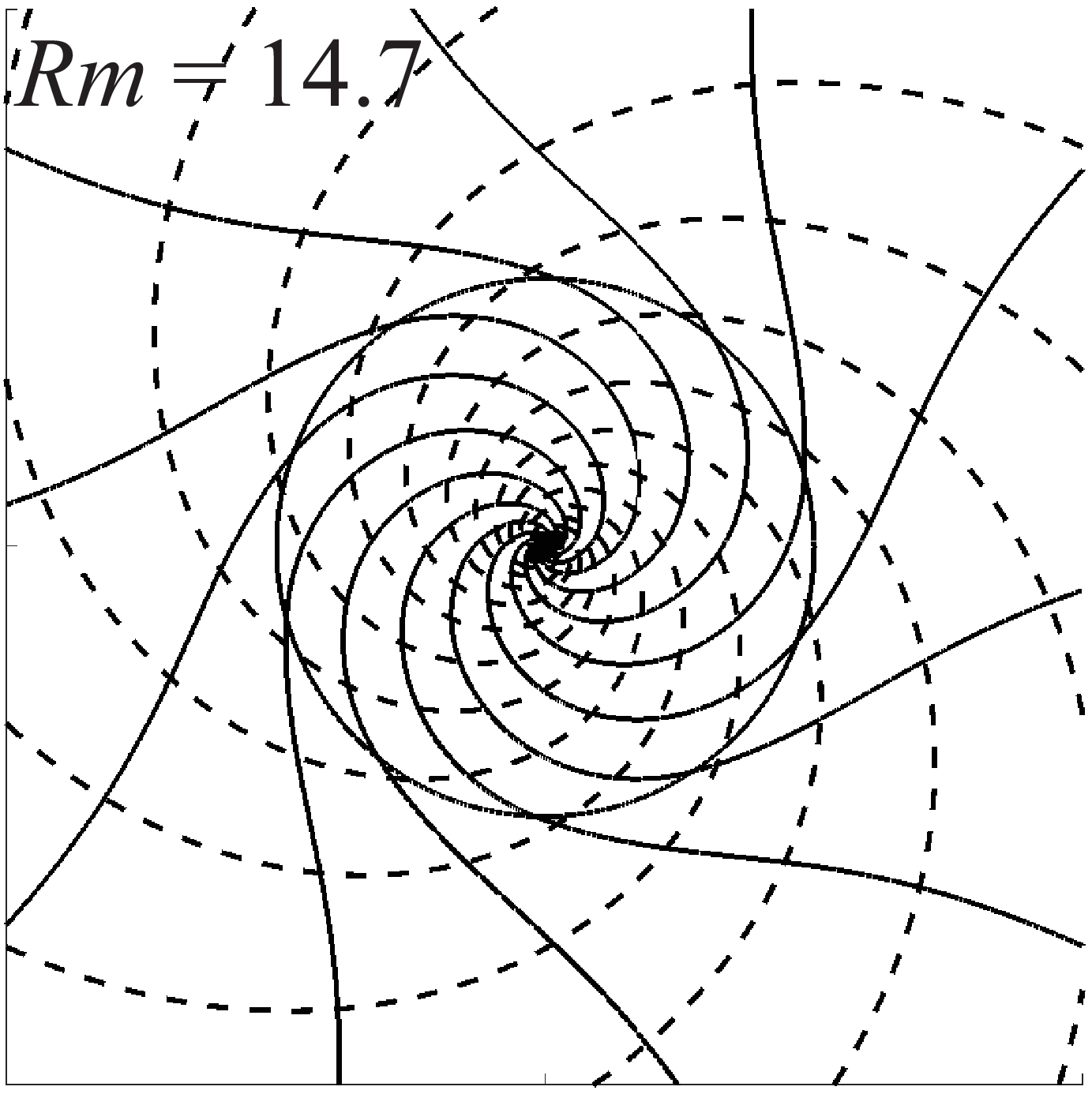}
\includegraphics[scale=0.22]{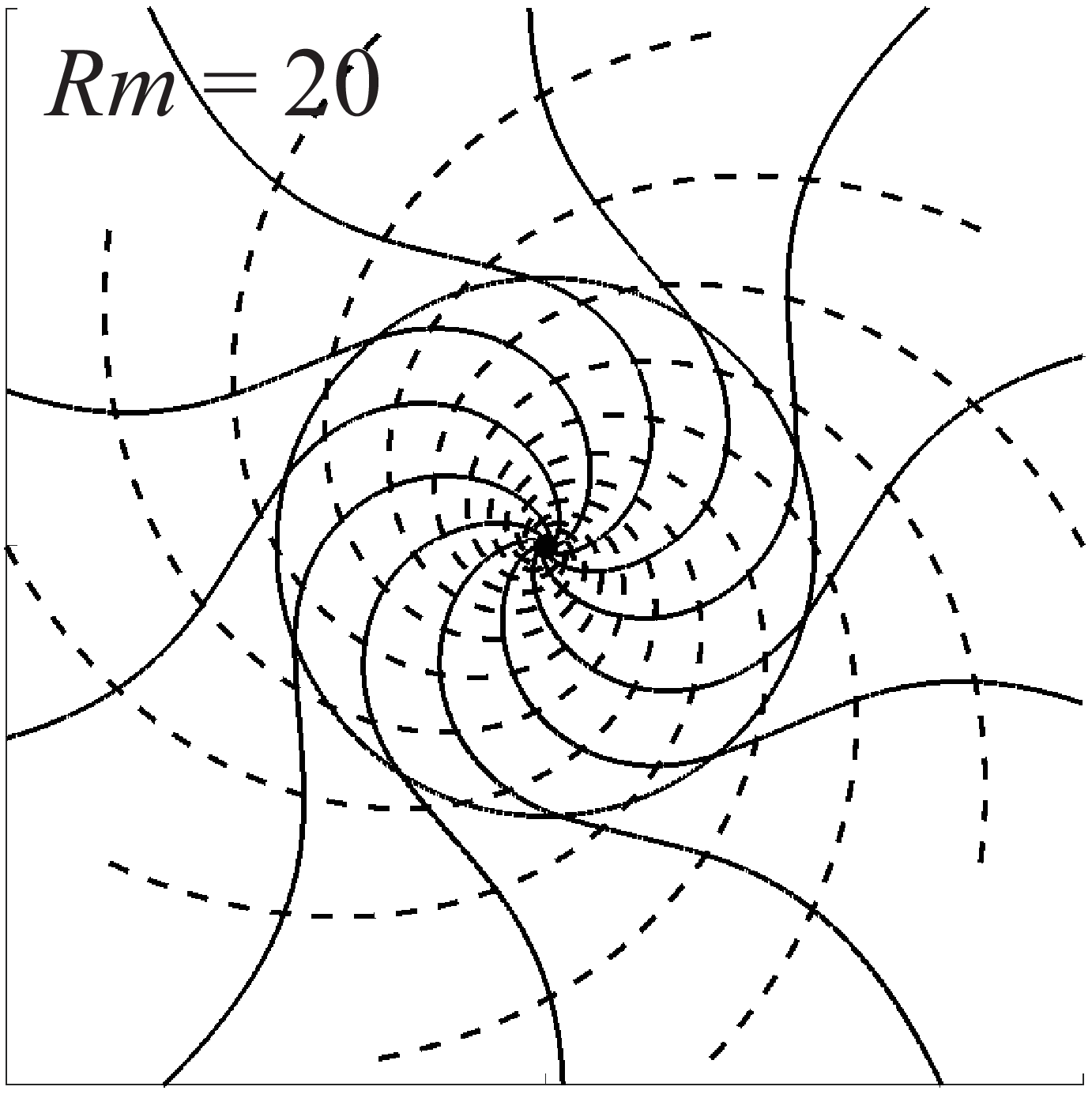}
\includegraphics[scale=0.22]{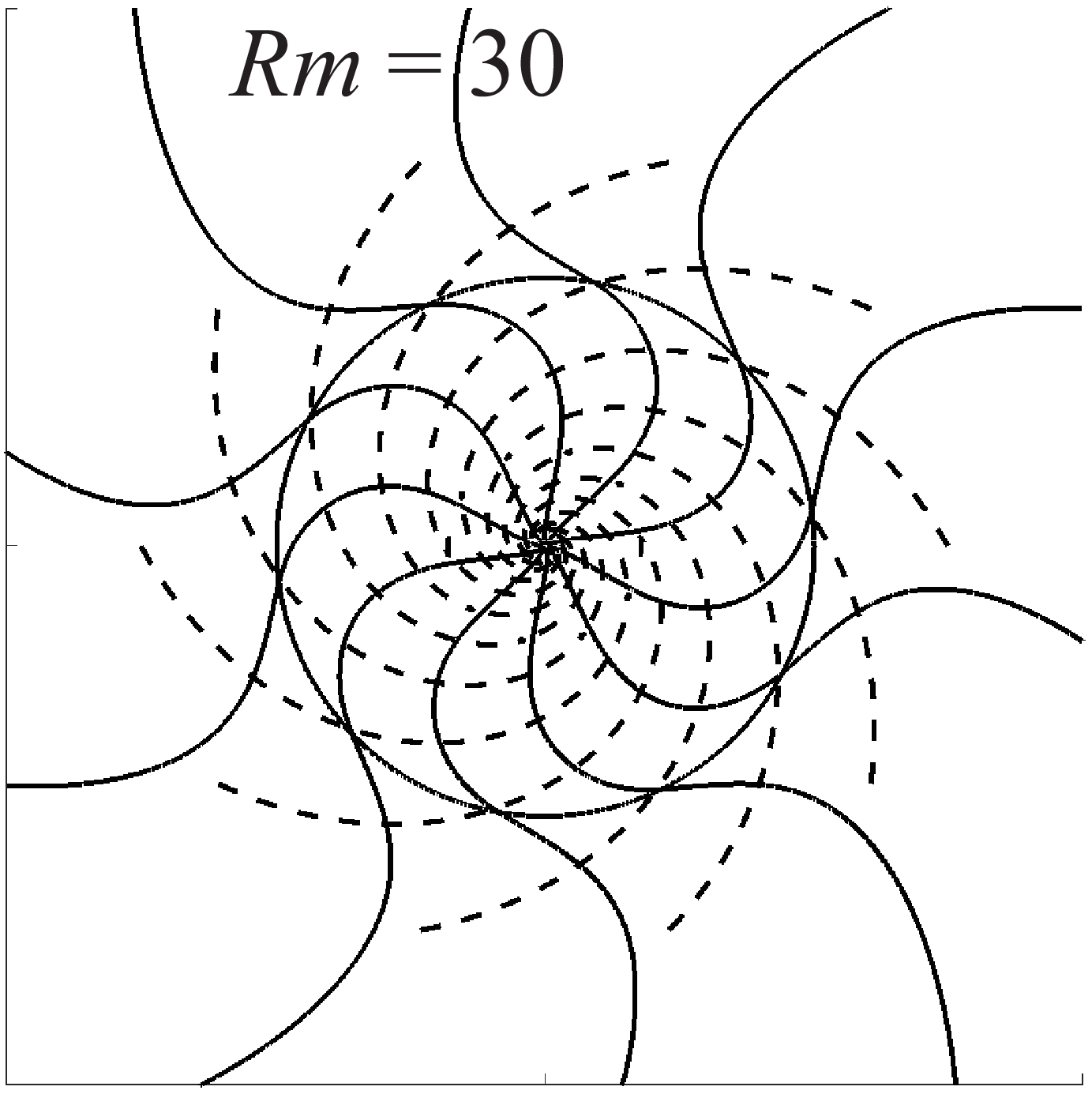}
\includegraphics[scale=0.22]{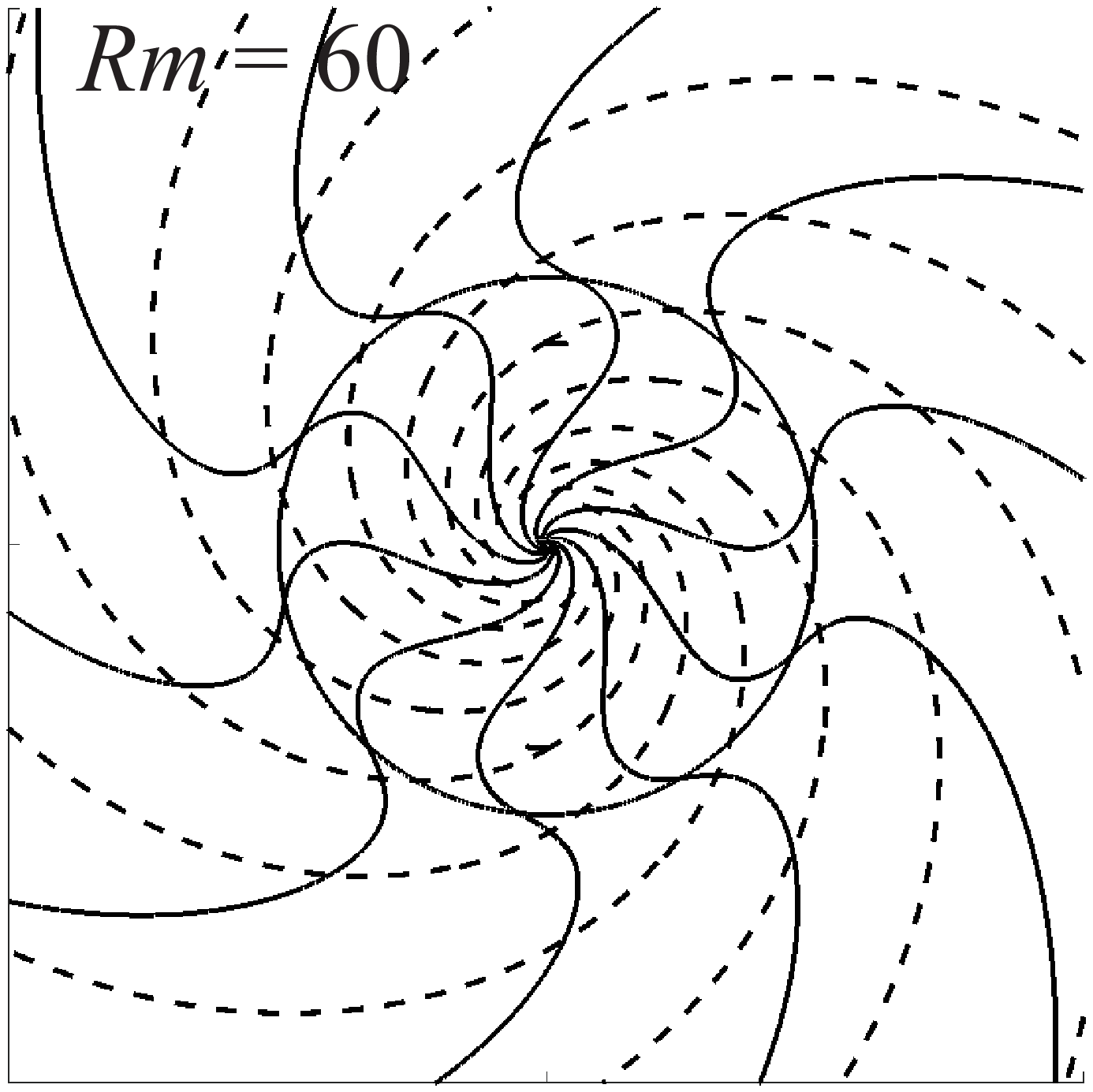}
\includegraphics[scale=0.22]{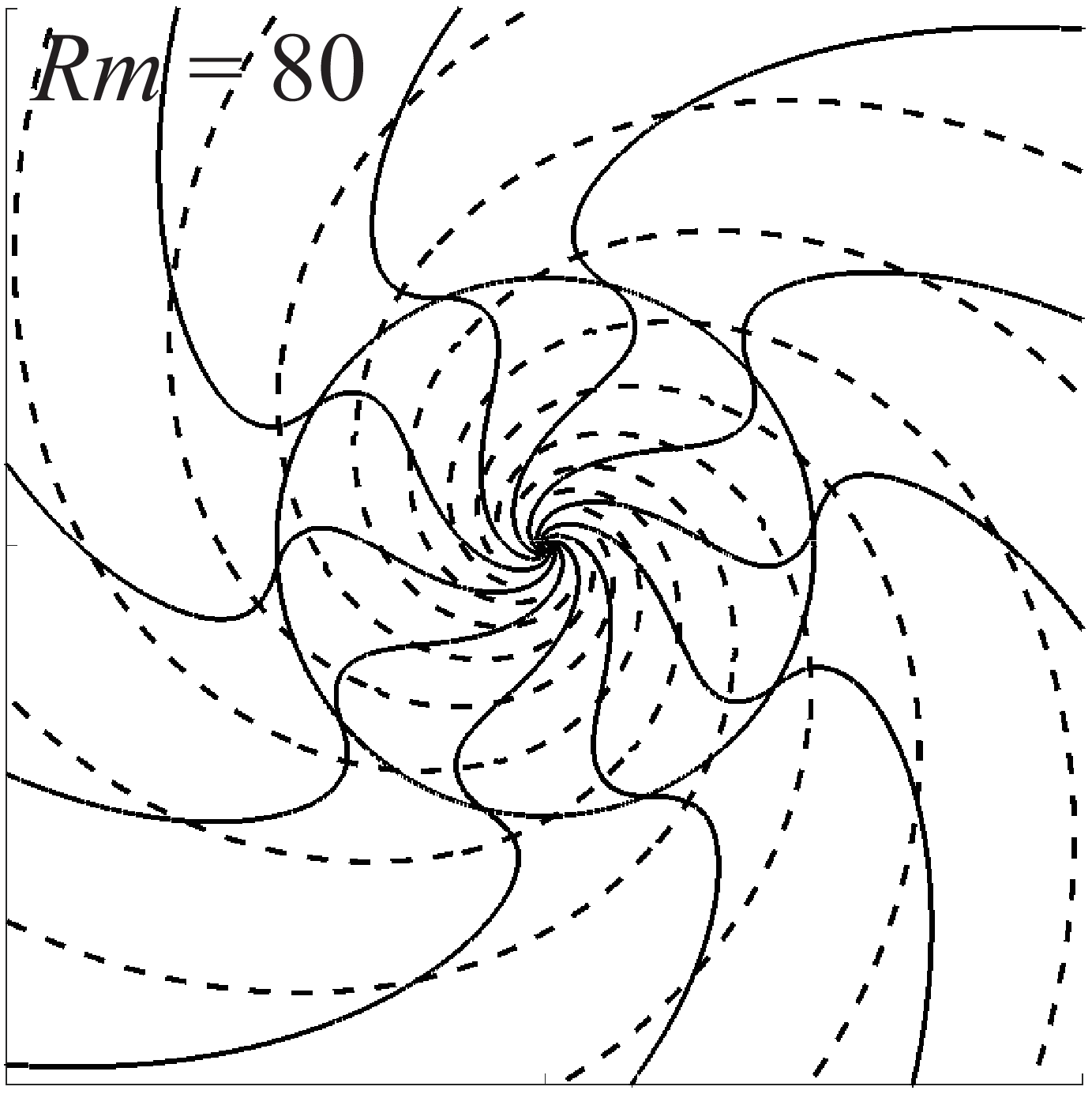}
\includegraphics[scale=0.22]{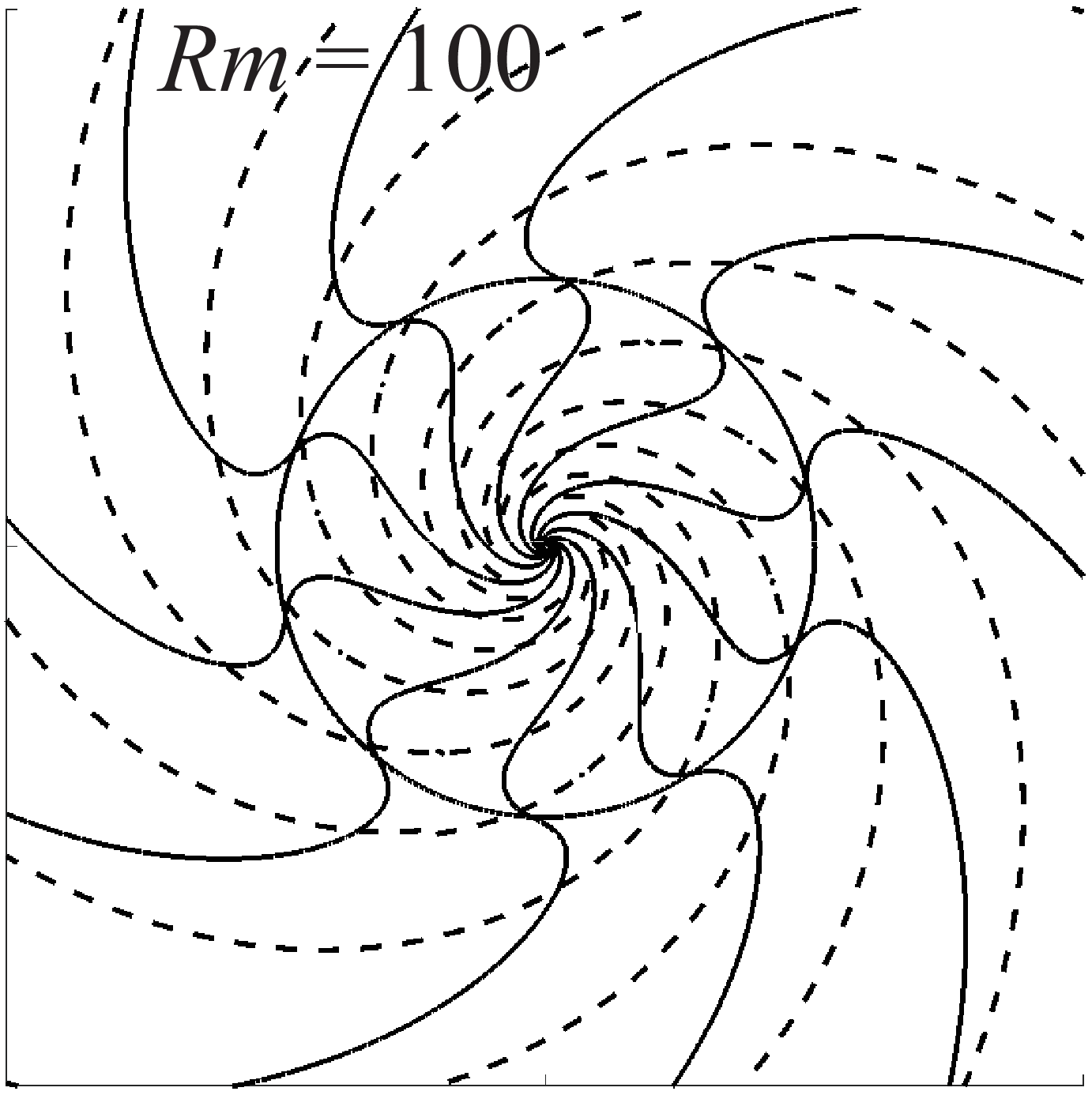}
\includegraphics[scale=0.22]{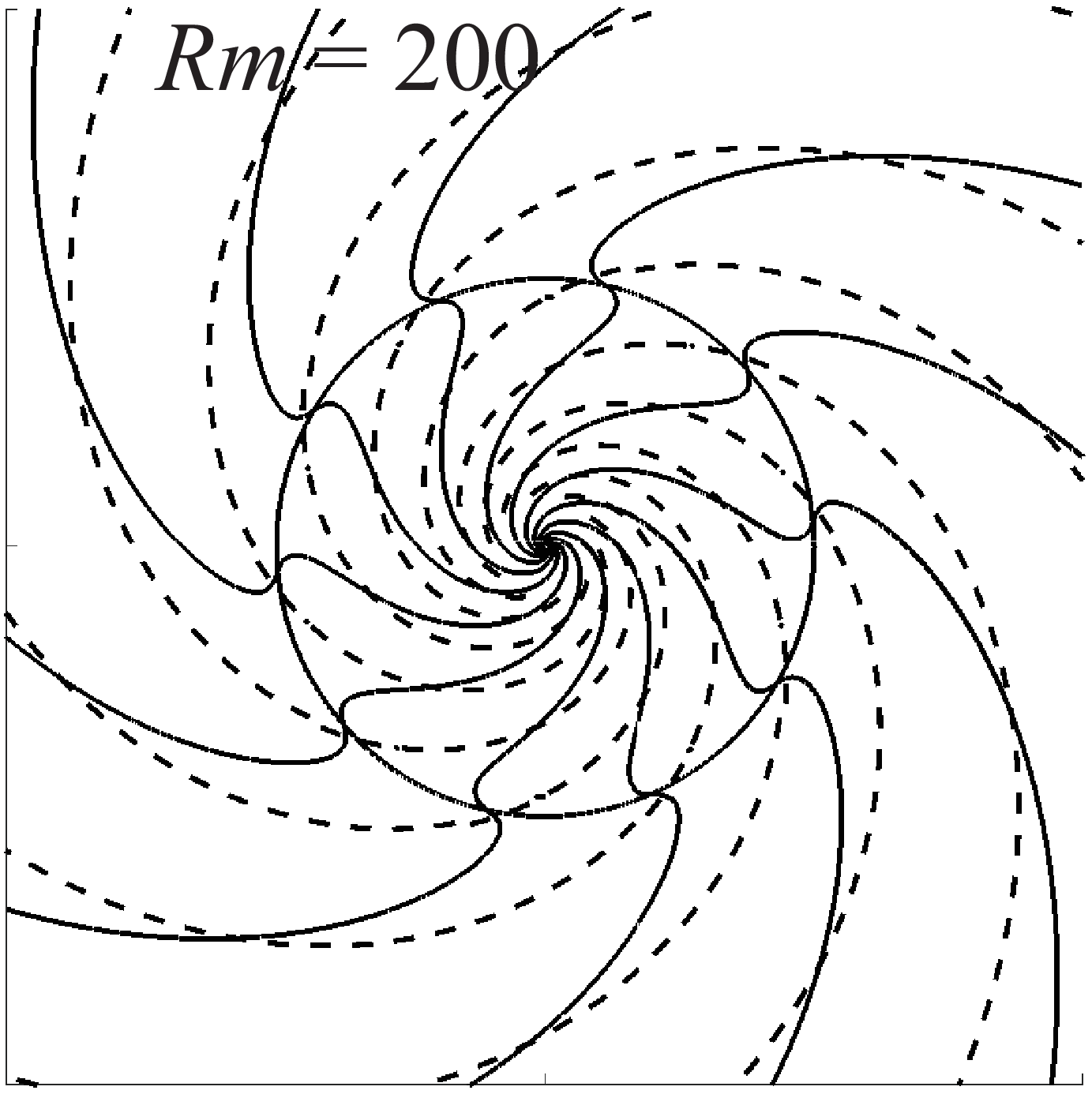}
\includegraphics[scale=0.22]{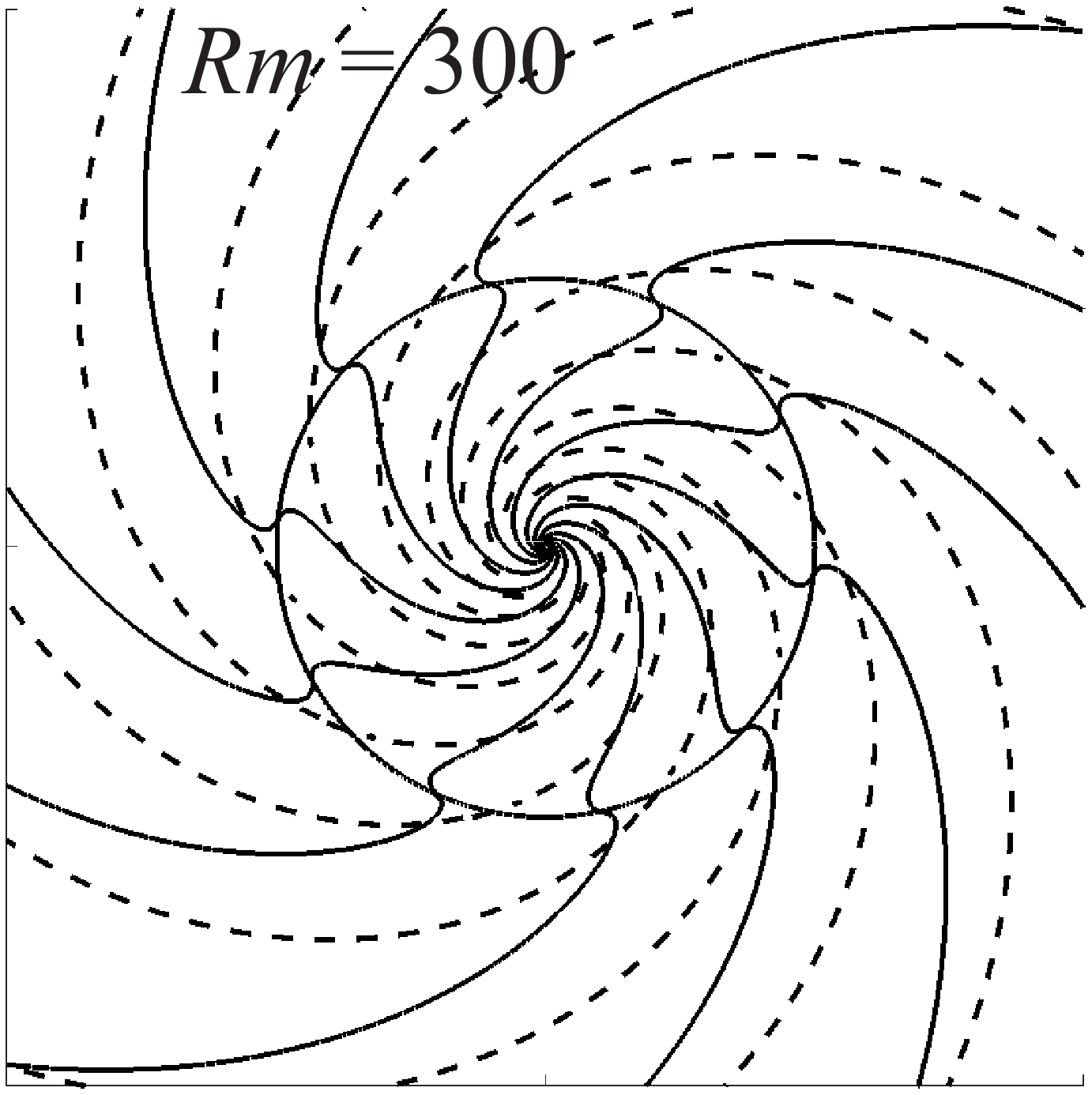}
\caption{Plots of the magnetic field lines (full) and the electric current lines (dashed) in the horizontal plane $(x,y)$, for $\mu=0$, $\sigma=10^6$, $\alpha=0.16 \pi$, and different values of $\Rm$.}
\label{fig:currentlines}
\end{center}
\end{figure}

\begin{figure}
\begin{center}
\includegraphics[scale=0.15]{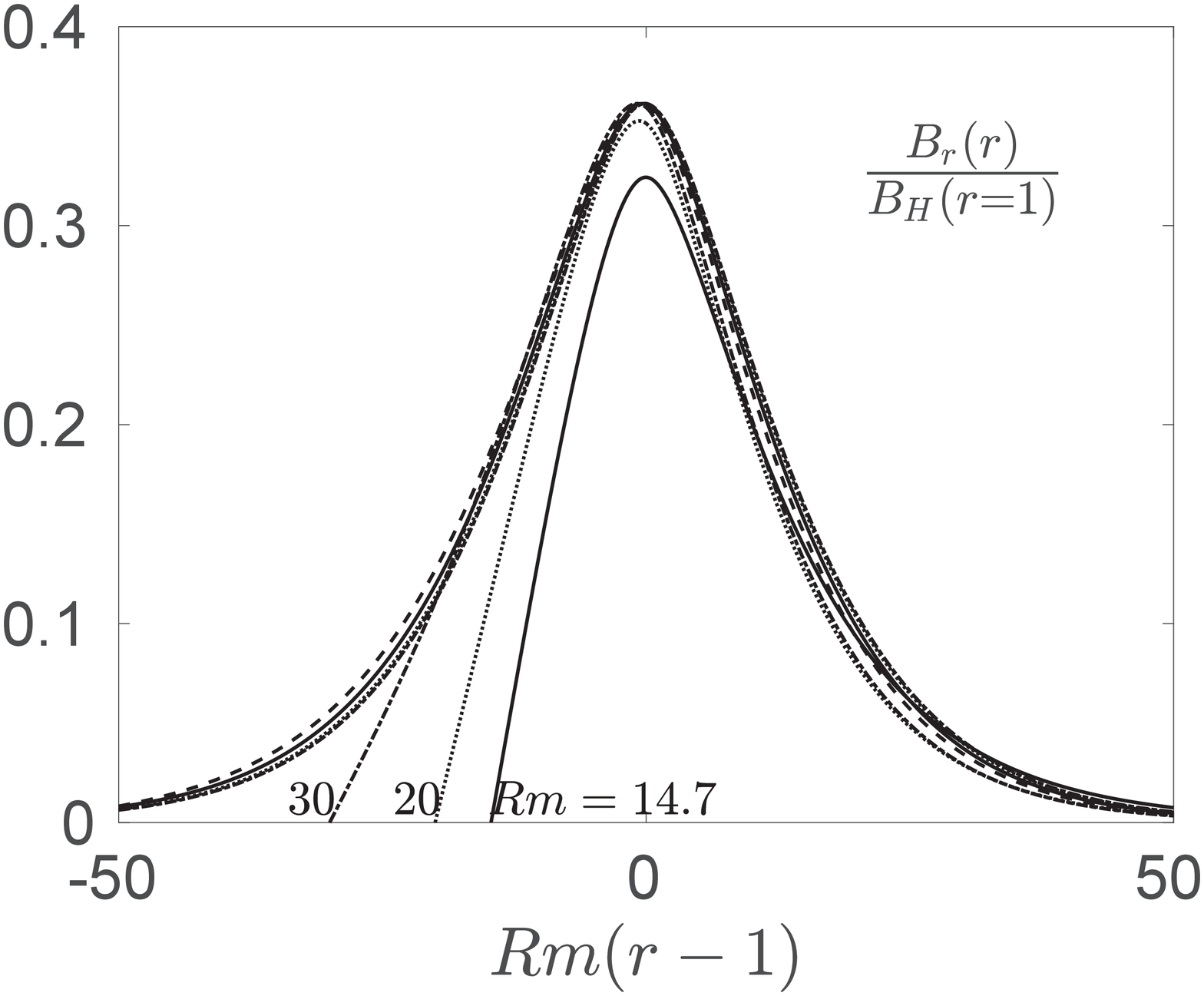}
\includegraphics[scale=0.15]{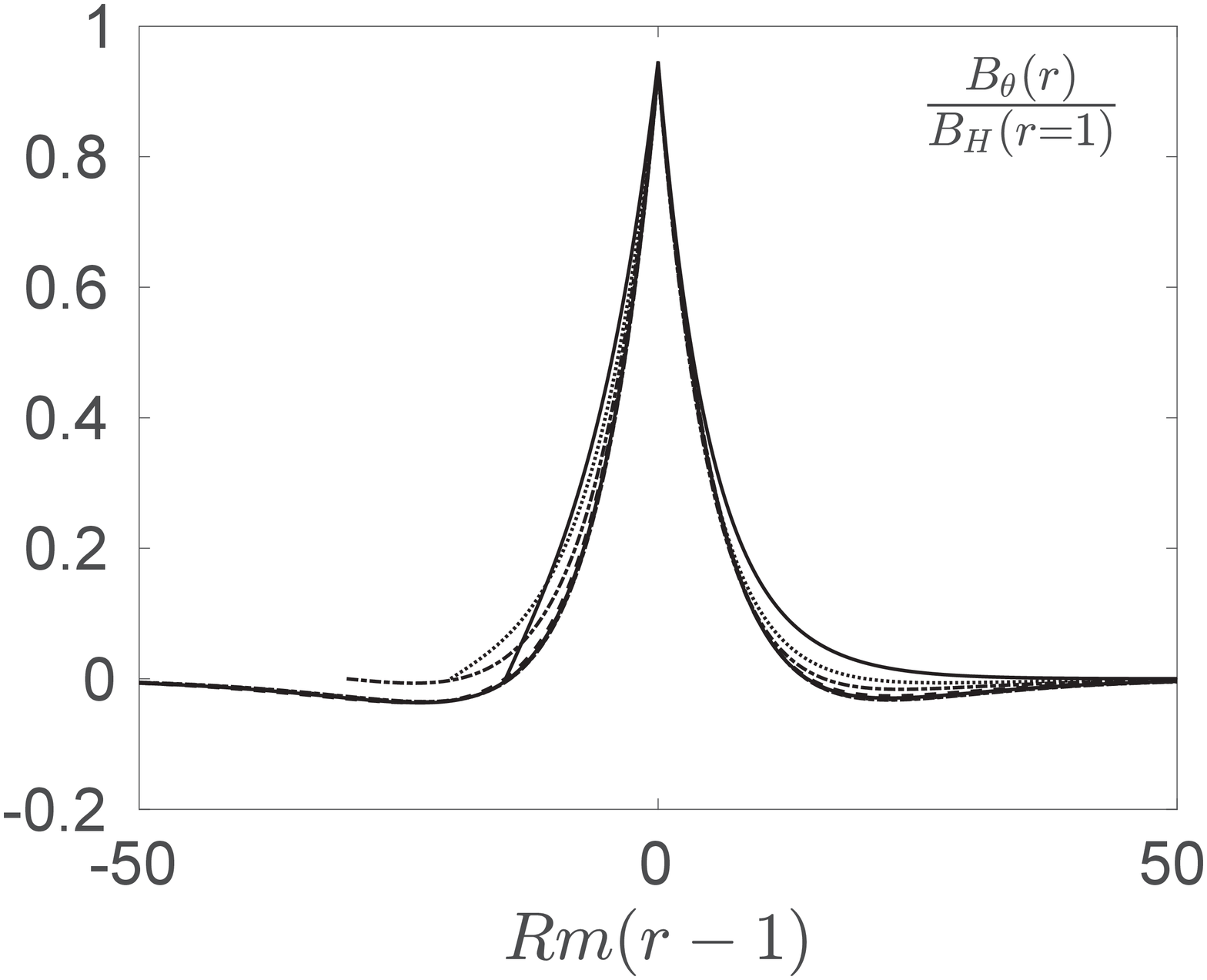}
\includegraphics[scale=0.15]{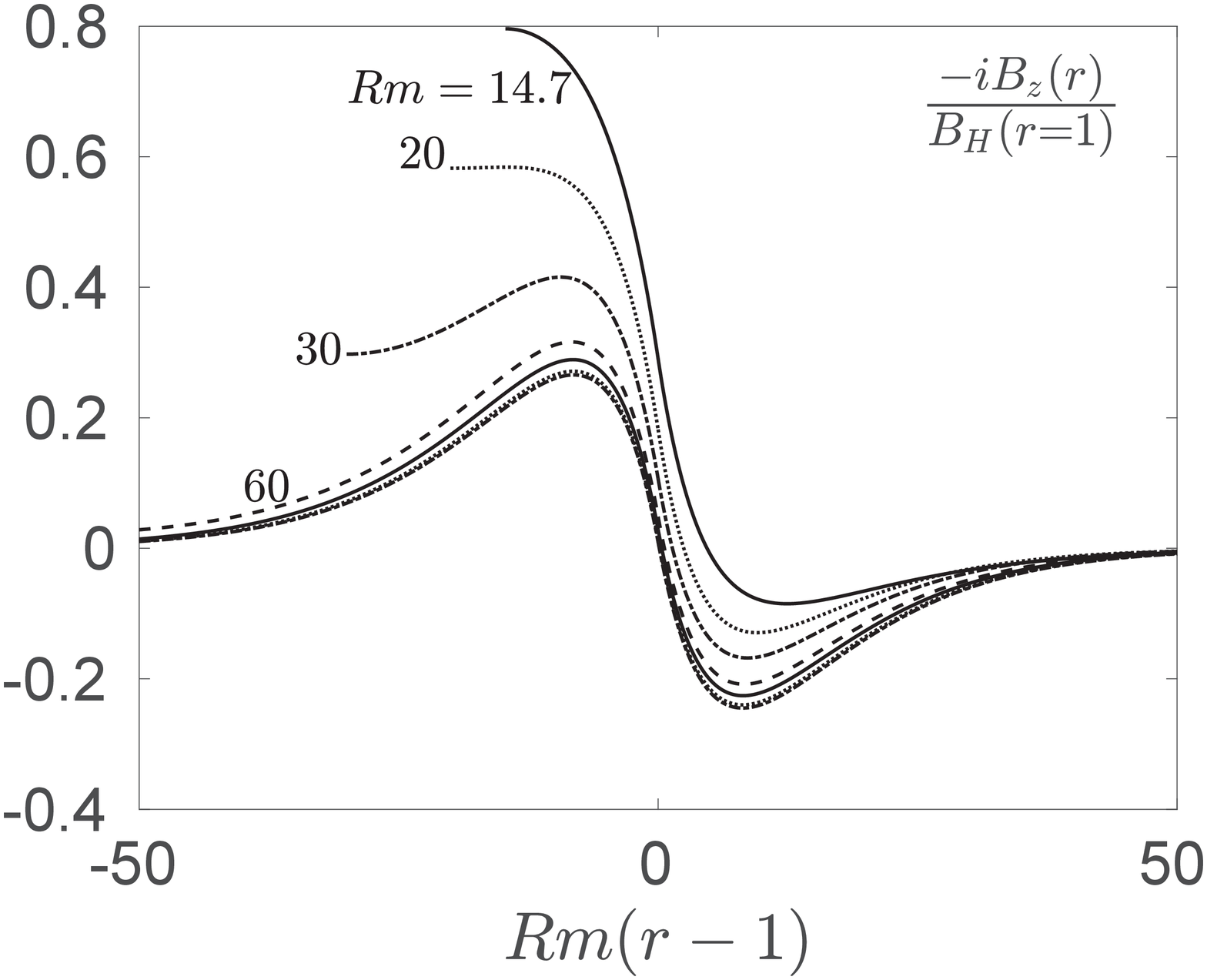}
\caption{From left to right, plots of $B_r$, $B_{\theta}$ and $-\ii B_z$, normalized by $B_H(r=1)$, versus $\Rm(r-1)$, for $\mu=0$, $\sigma=10^6$, $\alpha=0.16 \pi$ and $\Rm \in \{ 14.7,20,30,60,100,200,300\}$.}
\label{fig:BHBz}
\end{center}
\end{figure}

\begin{figure}
\begin{center}
\includegraphics[scale=0.15]{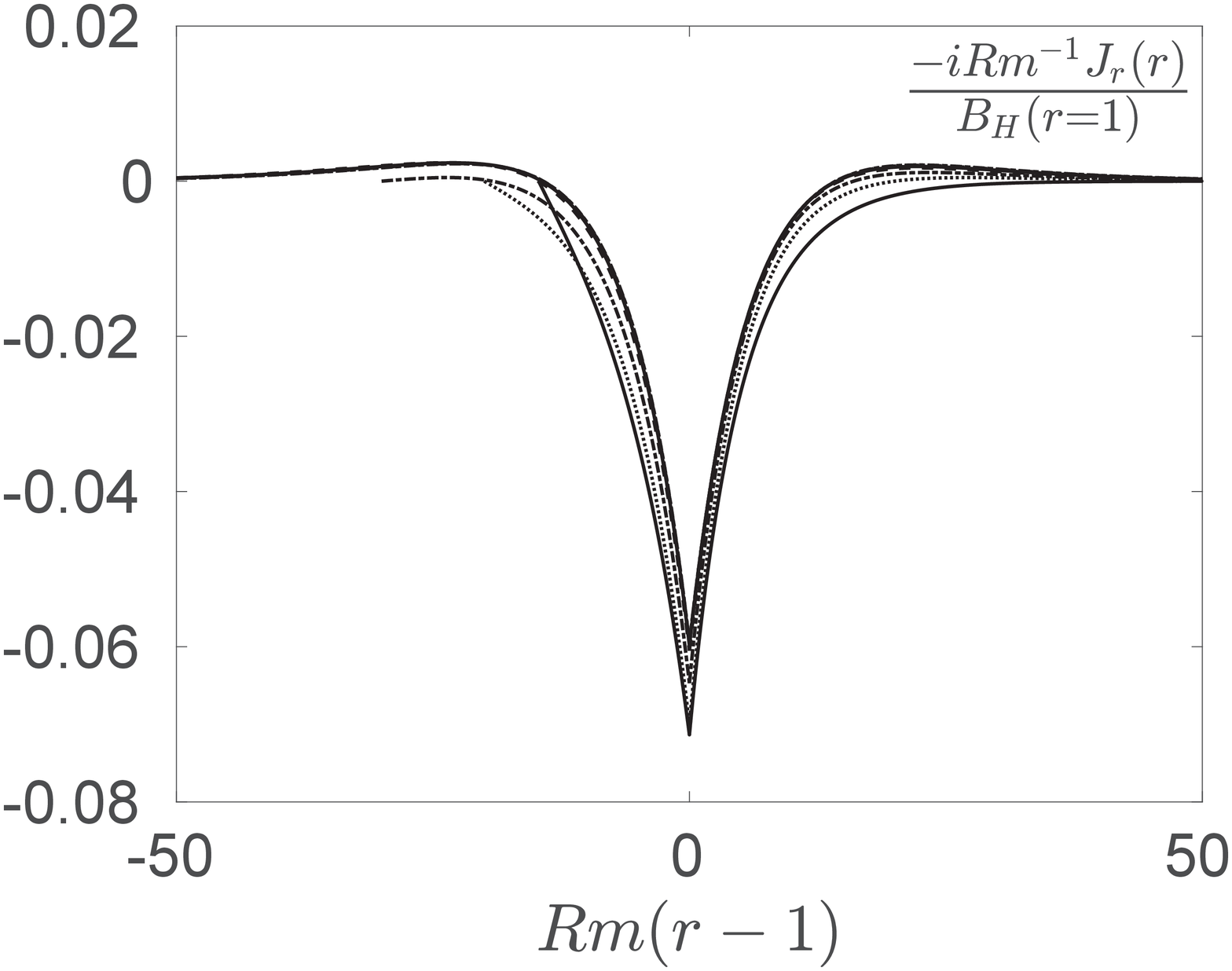}
\includegraphics[scale=0.15]{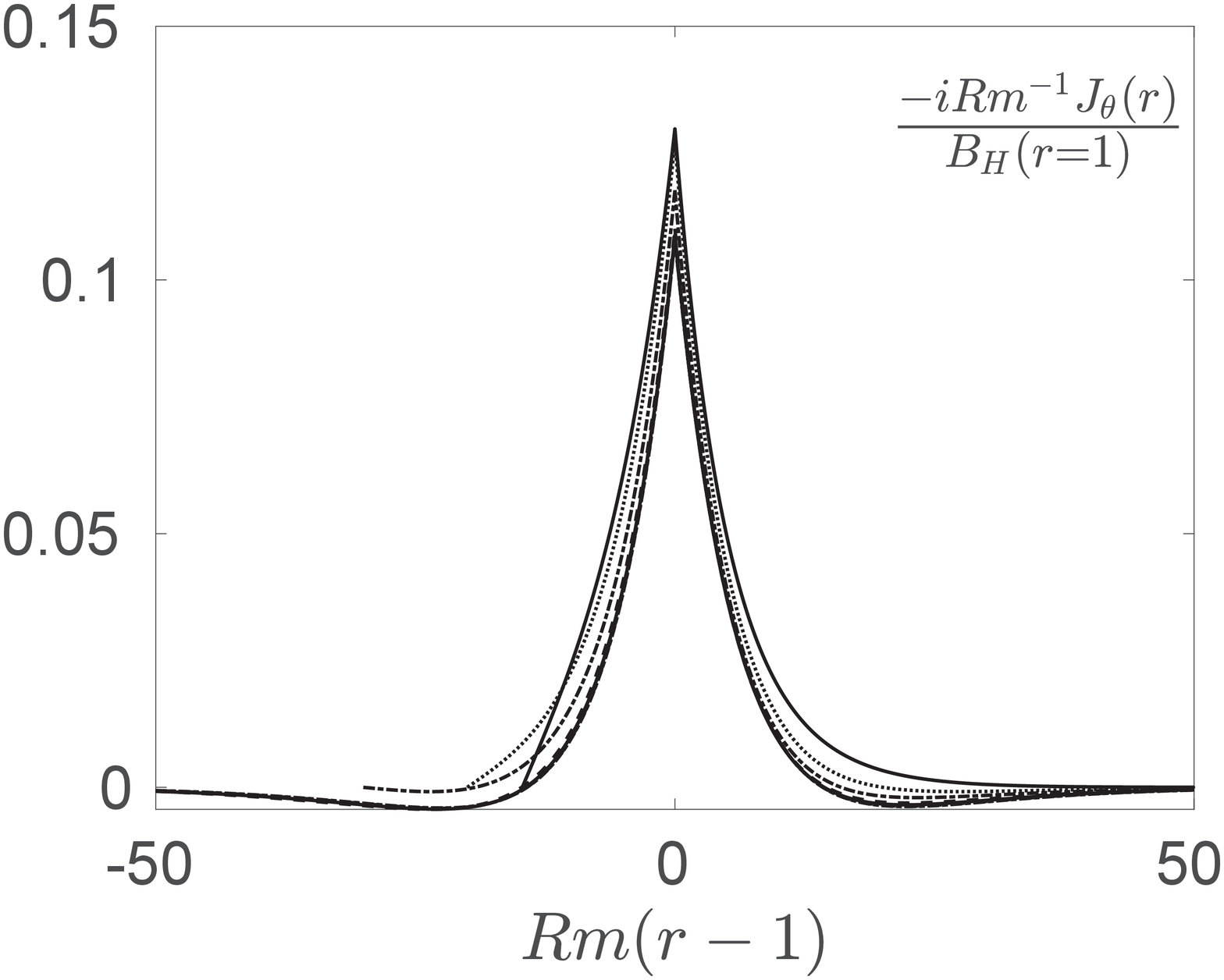}
\includegraphics[scale=0.15]{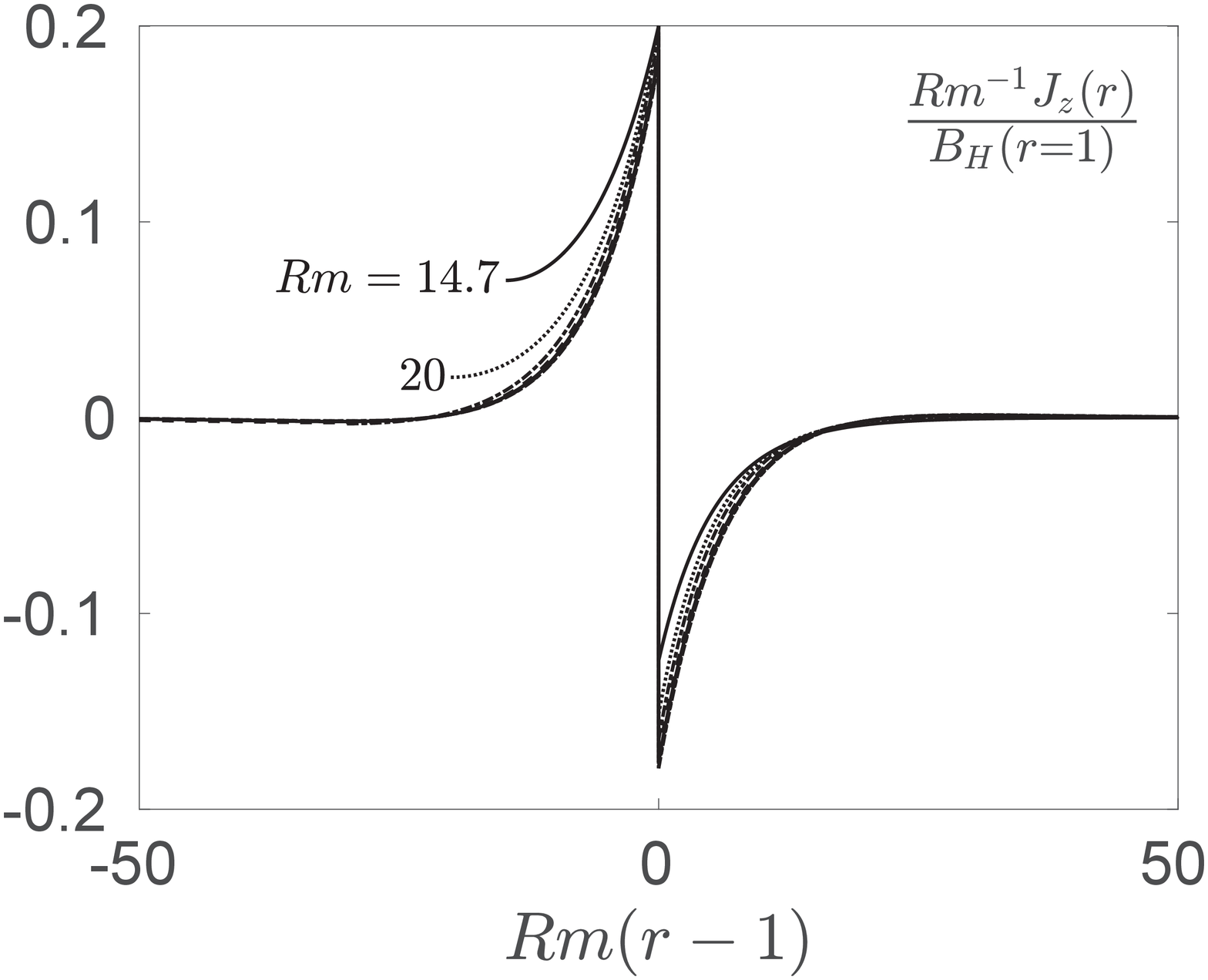}
\caption{From left to right, plots of $-\ii J_r$, $-\ii J_{\theta}$ and $J_z$, normalized by $\Rm\; B_H(r=1)$, versus $\Rm(r-1)$, for $\mu=0$, $\sigma=10^6$, $\alpha=0.16 \pi$ and $\Rm \in \{ 14.7,20,30,60,100,200,300\}$.}
\label{fig:JHJz}
\end{center}
\end{figure}

It is instructive to plot the geometries of the magnetic field and the current density for different values of $\Rm$. For that we use the expressions derived in (\ref{eq:Brlambda}-\ref{eq:Btetalambda}) and (\ref{eq:Bzlambda}) for the magnetic field, and (\ref{eq:Jrinf1}-\ref{eq:Jzsup1}) for the electrical current density. The geometries of the horizonthal magnetic field lines and electric current lines are plotted in Figure \ref{fig:currentlines}. 
The three components of the magnetic field normalized by $B_H(r=1)$ where $B_H$ is the modulus of the horizonthal component $B_H=\sqrt{B_r^2+B_{\theta}^2}$, are given in Figure \ref{fig:BHBz}.
The three components of the current density, normalized by $\Rm B_H(r=1)$, are given in Figure \ref{fig:JHJz}.

In Figure \ref{fig:currentlines}, we note that the geometry of the electric current lines in the $(x,y)$-plane do not depend on $\Rm$. This is due to the fact that $\sigma \gg 1$. Indeed, from the expressions of $J_r$ and $J_{\theta}$ given in Appendix \ref{sec:derivation of j} (\ref{eq:Jrinf1}-\ref{eq:Jtetasup1}) it can be shown that, for $\sigma \gg 1$ and provided that $\tilde{\gamma}/k^2$ is bounded, we have $J_{\theta}=-\frac{c}{s}J_r$ for both $r<1$ and $r>1$. This corresponds to have $\bJ \cdot \bq = 0$ or, equivalently, the electric current lines perpendicular to $\bq$. In contrast, the geometry of the horizonthal magnetic fields lines vary with $\Rm$, in such a way that a magnetic boundary layer seems to appear at $r=1$ for $\Rm \gg 1$. To quantify such a boundary layer, in Figures \ref{fig:BHBz} and  \ref{fig:JHJz} the components of $\bB$ and $\bJ$ are plotted versus $\Rm(r-1)$, and it is obvious that the curves merge as $\Rm$ increases, suggesting that the thickness of the boundary layer is of the order $O(\Rm^{-1})$.

\section{Conclusions}
We have shown that the anisotropic dynamo in cylindrical geometry is fast if the differential rotation is smooth, and furious if the shear is infinite.
In both cases the underlying mechanism is based on the stretching by differential rotation and anisotropic diffusion. For a smooth velocity profile, the dynamo occurs on a time scale equal to the turnover time $t_{\mathrm{Fast}}=|\Omega(a)|^{-1}$, where $a$ is some characteristic radius, indicating for example the one at which the shear is maximum. If the shear is infinite at $r=a$, the dynamo occurs on a time scale $t_{\mathrm{Furious}}$ even shorter that $|\Omega(a)|^{-1}$. In dimensional units, it is equal to
\begin{equation}
t_{\mathrm{Furious}}= (\Rm\;\Omega (a))^{-1}=(\sigma_{\perp}\mu_{\perp}a^2\Omega(a)^2)^{-1}. \label{eq:tau}
\end{equation}
For  $\Rm\gg1$ we then have $t_{\mathrm{Furious}} \ll t_{\mathrm{Fast}}$.
This new characteristic time $t_{\mathrm{Furious}}$ arises because, in the anisotropic dynamo, the mechanism of magnetic generation due to the anisotropic diffusion is particularly efficient, at least more efficient than the mechanism due to the isotropic diffusion on which, for example, the Ponomarenko dynamo is partially working on. As a result, the magnetic boundary layer in which the generation of the magnetic field occurs is thinner, leading to a magnetic growth faster than the fast Ponomarenko dynamo. To illustrate this, it is instructive to rewrite the system of equations (\ref{eq:gammaBrBteta}$\mathit{a,}\mathit{b}$) in the following schematical way
\begin{subequations}\label{eq:gammaBrBtetaanis}
\begin{align}
\gamma B_r &\sim - k^2 \Rm^{-1} B_r + k^2 \Rm^{-1} B_{\theta} , \label{eq:gammaBranis}\\
\gamma B_{\theta}  &\sim - k^2 \Rm^{-1} B_{\theta} + k^2 \Rm^{-1} B_r + r \Omega'(r) B_r \label{eq:gammaBtetaanis},
\end{align}
\end{subequations}
in which each term is a simplified expression of the original terms of (\ref{eq:gammaBrBteta}$\mathit{a,}\mathit{b}$).
On the right-hand side of (\ref{eq:gammaBrBtetaanis}$\mathit{a,}$$\mathit{b}$), the first term of each equation corresponds to the magnetic dissipation, acting against the dynamo action, the second tems are source terms for the dynamo due to the anisotropic diffusion, while the third term of (\ref{eq:gammaBrBtetaanis}$\mathit{a}$) is also a source term, due to the velocity shear.
Leaving aside temporarily the velocity term $r\Omega'(r) B_r$, all other terms of (\ref{eq:gammaBrBtetaanis}$\mathit{a,}\mathit{b}$) are of the same order of magnitude provided that
\begin{equation}
 B_r\sim B_{\theta}, \;\;\; \text{and} \;\;\; \gamma \sim k^2 \Rm^{-1}. 
\end{equation}
The velocity term can be estimated as $r\Omega'(r) \sim 1$ if the shear is smooth and $r\Omega'(r) \sim k$ if the shear is infinite.
Assuming that in (\ref{eq:gammaBrBtetaanis}$\mathit{a,}\mathit{b}$) the velocity term is also of the same order of magnitude as the other terms, leads to $\gamma \sim 1$ and $k\sim Rm^{\frac{1}{2}}$ for the smooth shear, and to $\gamma\sim k\sim\Rm$ for the infinite shear. The thickness of the magnetic boundary layer that can be estimated as $\delta \sim k^{-1}$, then scales as $\delta \sim Rm^{-\frac{1}{2}}$ in the smooth case, and $\delta \sim Rm^{-1}$ in the infinite shear case. 
These orders of magnitude confirmed by our previous findings, clearly establish the crucial role of the boundary layers.  

In order to capture the difference with the Ponomarenko dynamo we can rewrite, again schematically, the system of equations given in \citet{Gilbert1988,Gilbert2003} as 
\begin{subequations}\label{eq:gammaBrBtetapono}
\begin{align}
\gamma B_r &\sim - k^2 \Rm^{-1} B_r + k \Rm^{-1} B_{\theta} , \label{eq:gammaBrpono}\\
\gamma B_{\theta}  &\sim - k^2 \Rm^{-1} B_{\theta} + k \Rm^{-1} B_r + r \Omega'(r) B_r \label{eq:gammaBtetapono}.
\end{align}
\end{subequations}
In  (\ref{eq:gammaBrBtetapono}$\mathit{a}$),  $k \Rm^{-1} B_{\theta}$ is again a source term corresponding to the generation of $B_r$ from $B_{\theta}$, coming  from the isotropic diffusion of  $B_{\theta}$ in the $r$-direction. This term is to contrast with the anisotropic diffusion term $k^2 \Rm^{-1} B_{\theta}$ in (\ref{eq:gammaBrBtetaanis}$\mathit{a}$). In both cases the diffusion is involved, but they have different orders of magnitude for $k\gg1$.
Assuming that the terms on the right-hand side of (\ref{eq:gammaBrBtetapono}$\mathit{a}$) are of the same order of magnitude leads to 
\begin{equation}
 B_{\theta}\sim k B_r, \;\;\; \text{and} \;\;\; \gamma \sim k^2 \Rm^{-1}. 
\end{equation}
Taking the same estimation of the velocity term $r\Omega'(r)$ as above, $r\Omega'(r) \sim 1$ if the shear is smooth and $\Omega'(r) \sim k$ if the shear is infinite, and assuming that it is of the same order of magnitude as the other terms of (\ref{eq:gammaBrBtetapono}$\mathit{b}$), except 
the term $k \Rm^{-1} B_r$ which is smaller for $k\gg1$, leads to $\gamma \sim k^{-1}\sim Rm^{-\frac{1}{3}}$ for the smooth shear, and to $\gamma\sim 1$ and $ k\sim \Rm^{\frac{1}{2}}$ for the infinite shear. The thickness of the magnetic boundary layer, $\delta \sim k^{-1}$, then scales as $\delta \sim Rm^{-\frac{1}{3}}$ in the smooth case, and $\delta \sim Rm^{-\frac{1}{2}}$ in the infinite shear case. 

Eventually, a unique characteristic time can be defined, as
\begin{equation}
 t = \sigma_{\perp} \mu_{\perp} \delta^2,
\label{eq:characteristic time}
\end{equation}
with,  for the Ponomarenko dynamo, $\sigma_{\perp}=\sigma_{\parallel}$ and $\mu_{\perp}=\mu_{\parallel}$.
In (\ref{eq:characteristic time}), $t$ corresponds to the magnetic diffusion time through a magnetic boundary layer of thickness $\delta$.
Replacing $\delta/a$ by either $Rm^{-\frac{1}{3}}$, $Rm^{-\frac{1}{2}}$ or $Rm^{-1}$ leads to a characteristic time equal to  $Rm^{-\frac{1}{3}}\Omega(a)^{-1}$, $\Omega(a)^{-1}$ or $Rm^{-1}\Omega(a)^{-1}$, as summarized in Table \ref{tab:summary}.
\begin{table}
\begin{center}
\begin{tabular}{l c c c}
                      & Slow                         & Fast                                & Furious  \\
                      & (Ponomarenko)            & (Ponomarenko, Anisotropic) & (Anisotropic)\\
\\
$\delta/a$ 	   & $Rm^{-\frac{1}{3}}$ & $Rm^{-\frac{1}{2}}$       & $Rm^{-1}$\\
$t\Omega(a)$  & $Rm^{\frac{1}{3}}$  & 1                                    & $Rm^{-1}$\\
\end{tabular}
\label{tab:summary}
\end{center}
\caption{$\Rm$-power scalings  of the thickness $\delta$ of the magnetic boundary layer, and of the characteristic time $t$ of the dynamo. These two quantities satisfy the unique formula (\ref{eq:characteristic time}).}
\end{table}
Therefore, as mentioned above, it is mainly the thickness of the boundary layer that governs the characteristic time of the dynamo action, and thus the ability to have a slow, fast or furious dynamo for increasingly thin boundary layers. 

Such an anisotropic dynamo can be designed at the laboratory scale, using appropriate conducting layers or coils, or high-permeability materials, in order to mimic the homogeneous anisotropy considered here. According to figure \ref{fig:asymptotics}, to successfully demonstrate the furious aspect of such a dynamo, a minimum magnetic Reynolds number of about 30 would be necessary, which is about twice more than the value 14.6 of the dynamo threshold. From the estimate given in \citet{Plunian2020}, assuming an electrical conductivity equal to the one of copper would require an inner cylinder with a radius of 0.05 m and a rotation frequency of about 25 Hz, which is feasible in the laboratory. In natural objects where the magnetic Reynolds number is much larger, fast or furious dynamo action should be favoured, provided that the anisotropy of the electrical conductivity does play a role, as can be expected for example in spiral arms galaxies where $\Rm \gg 1$ and for which our anisotropic model might be a good approximation.

\section*{Acknowledgements}
We acknowledge Andrew Gilbert for his interest in our work and his helpful comments.\\
\\
Declaration of Interests. The authors report no conflict of interest.
 
\bibliographystyle{jfm}

\begin{thebibliography}{33}
\expandafter\ifx\csname natexlab\endcsname\relax\def\natexlab#1{#1}\fi
\def\au#1{#1} \def\ed#1{#1} \def\yr#1{#1}\def\at#1{#1}\def\jt#1{\textit{#1}}
  \def\bt#1{#1}\def\bvol#1{\textbf{#1}} \def\vol#1{#1} \def\pg#1{#1}
  \def\publ#1{#1}\def\arxiv#1{#1}\def\org#1{#1}\def\st#1{\textit{#1}}

\bibitem[Abramowitz \& Stegun(1968)]{Abramowitz1968}
{\sc \au{Abramowitz, M.} \& \au{Stegun, I.~A.}} \yr{1968} {\em Handbook of
  Mathematical Functions with Formulas, Graphs and Mathemarical Tables\/}.
  \publ{Dover Publications, New York}.

\bibitem[Alboussi\`ere {\em et~al.\/}(2020)Alboussi\`ere, Drif \&
  Plunian]{Alboussiere2020}
{\sc \au{Alboussi\`ere, T.}, \au{Drif, K.} \& \au{Plunian, F.}} \yr{2020}
  \at{Dynamo action in sliding plates of anisotropic electrical conductivity}.
  \jt{Physical Review E}  \bvol{101},  \pg{033107}.

\bibitem[{Braginskii}(1965)]{Braginskii1965}
{\sc \au{{Braginskii}, S.I.}} \yr{1965}  \at{{Transport Processes in a
  Plasma}}.  \jt{Reviews of Plasma Physics}  \bvol{1},  \pg{205}.

\bibitem[Brandenburg \& Subramanian(2005)]{Brandenburg2005}
{\sc \au{Brandenburg, A.} \& \au{Subramanian, K.}} \yr{2005}  \at{Astrophysical
  magnetic fields and nonlinear dynamo theory}.  \jt{Physics Reports}
  \bvol{417}~(1),  \pg{1--209}.

\bibitem[Childress \& Gilbert(1995)]{Childress1995}
{\sc \au{Childress, S.} \& \au{Gilbert, A.~D.}} \yr{1995} {\em Stretch, Twist,
  Fold: The Fast Dynamo\/}.  \publ{Springer}.

\bibitem[Cowling(1934)]{Cowling1934}
{\sc \au{Cowling, T.~G.}} \yr{1934}  \at{The magnetic field of sunspots}.
  \jt{Mon. Not. R. Astr. Soc.}  \bvol{94},  \pg{39--48}.

\bibitem[Deuss(2014)]{Deuss2014}
{\sc \au{Deuss, A.}} \yr{2014}  \at{Heterogeneity and anisotropy of earth's
  inner core}.  \jt{Annual Review of Earth and Planetary Sciences}
  \bvol{42}~(1),  \pg{103--126}.

\bibitem[Favier \& Proctor(2013)]{Favier2013}
{\sc \au{Favier, B.} \& \au{Proctor, M. R.~E.}} \yr{2013}  \at{Growth rate
  degeneracies in kinematic dynamos}.  \jt{Phys. Rev. E}  \bvol{88},
  \pg{031001}.

\bibitem[Gallet {\em et~al.\/}(2012)Gallet, P{\'{e}}tr{\'{e}}lis \&
  Fauve]{Gallet2012}
{\sc \au{Gallet, B.}, \au{P{\'{e}}tr{\'{e}}lis, F.} \& \au{Fauve, S.}}
  \yr{2012}  \at{Dynamo action due to spatially dependent magnetic
  permeability}.  \jt{Europhysics Letters}  \bvol{97}~(6),  \pg{69001}.

\bibitem[Gallet {\em et~al.\/}(2013)Gallet, Pétrélis \& Fauve]{Gallet2013}
{\sc \au{Gallet, B.}, \au{Pétrélis, F.} \& \au{Fauve, S.}} \yr{2013}
  \at{Spatial variations of magnetic permeability as a source of dynamo
  action}.  \jt{Journal of Fluid Mechanics}  \bvol{727},  \pg{161–190}.

\bibitem[Gilbert(1988)]{Gilbert1988}
{\sc \au{Gilbert, Andrew~D.}} \yr{1988}  \at{Fast dynamo action in the
  {P}onomarenko dynamo}.  \jt{Geophysical \& Astrophysical Fluid Dynamics}
  \bvol{44}~(1-4),  \pg{241--258}.

\bibitem[Gilbert(2003)]{Gilbert2003}
{\sc \au{Gilbert, Andrew~D.}} \yr{2003}  \bt{ \at{Chapter 9 - {D}ynamo
  {T}heory}}.  \st{Handbook of Mathematical Fluid Dynamics},  \vol{vol.~2},
  \pg{pp. 355--441}.  \publ{North-Holland}.

\bibitem[Kreuzahler {\em et~al.\/}(2017)Kreuzahler, Ponty, Plihon, Homann \&
  Grauer]{Kreuzahler2017}
{\sc \au{Kreuzahler, S.}, \au{Ponty, Y.}, \au{Plihon, N.}, \au{Homann, H.} \&
  \au{Grauer, R.}} \yr{2017}  \at{Dynamo enhancement and mode selection
  triggered by high magnetic permeability}.  \jt{Physical Review Letters}
  \bvol{119},  \pg{234501}.

\bibitem[Lortz(1989)]{Lortz1989}
{\sc \au{Lortz, D.}} \yr{1989}  \at{Axisymmetric dynamo solutions}.  \jt{Z.
  Naturforsch}  \bvol{44a},  \pg{1041--1045}.

\bibitem[Lowes \& Wilkinson(1963)]{Lowes1963}
{\sc \au{Lowes, F.~J.} \& \au{Wilkinson, I.}} \yr{1963}  \at{Geomagnetic
  dynamo: A laboratory model}.  \jt{Nature}  \bvol{198},  \pg{1158--1160}.

\bibitem[Lowes \& Wilkinson(1968)]{Lowes1968}
{\sc \au{Lowes, F.~J.} \& \au{Wilkinson, I.}} \yr{1968}  \at{Geomagnetic
  dynamo: An improved laboratory model}.  \jt{Nature}  \bvol{219},
  \pg{717--718}.

\bibitem[Marcotte {\em et~al.\/}(2021)Marcotte, Gallet, P\'etr\'elis \&
  Gissinger]{Marcotte2021}
{\sc \au{Marcotte, F.}, \au{Gallet, B.}, \au{P\'etr\'elis, F.} \&
  \au{Gissinger, C.}} \yr{2021}  \at{Enhanced dynamo growth in nonhomogeneous
  conducting fluids}.  \jt{Phys. Rev. E}  \bvol{104},  \pg{015110}.

\bibitem[Miralles {\em et~al.\/}(2013)Miralles, Bonnefoy, Bourgoin, Odier,
  Pinton, Plihon, Verhille, Boisson, Daviaud \& Dubrulle]{Miralles2013}
{\sc \au{Miralles, S.}, \au{Bonnefoy, N.}, \au{Bourgoin, M.}, \au{Odier, P.},
  \au{Pinton, J.-F.}, \au{Plihon, N.}, \au{Verhille, G.}, \au{Boisson, J.},
  \au{Daviaud, F.} \& \au{Dubrulle, B.}} \yr{2013}  \at{Dynamo threshold
  detection in the von {K}\'arm\'an sodium experiment}.  \jt{Physical Review E}
   \bvol{88},  \pg{013002}.

\bibitem[Nore {\em et~al.\/}(2018)Nore, Castanon~Quiroz, Cappanera \&
  Guermond]{Nore2018}
{\sc \au{Nore, C.}, \au{Castanon~Quiroz, D.}, \au{Cappanera, L.} \&
  \au{Guermond, J.-L.}} \yr{2018}  \at{Numerical simulation of the von
  {K}ármán sodium dynamo experiment}.  \jt{Journal of Fluid Mechanics}
  \bvol{854},  \pg{164–195}.

\bibitem[Ohta {\em et~al.\/}(2018)Ohta, Nishihara, Sato, Hirose, Yagi,
  Kawaguchi, Hirao \& Ohishi]{Ohta2018}
{\sc \au{Ohta, K.}, \au{Nishihara, Y.}, \au{Sato, Y.}, \au{Hirose, K.},
  \au{Yagi, T.}, \au{Kawaguchi, S.~I.}, \au{Hirao, N.} \& \au{Ohishi, Y.}}
  \yr{2018}  \at{An experimental examination of thermal conductivity anisotropy
  in hcp iron}.  \jt{Frontiers in Earth Science}  \bvol{6},  \pg{176}.

\bibitem[P\'etr\'elis {\em et~al.\/}(2016)P\'etr\'elis, Alexakis \&
  Gissinger]{Petrelis2016}
{\sc \au{P\'etr\'elis, F.}, \au{Alexakis, A.} \& \au{Gissinger, C.}} \yr{2016}
  \at{Fluctuations of electrical conductivity: A new source for astrophysical
  magnetic fields}.  \jt{Phys. Rev. Lett.}  \bvol{116},  \pg{161102}.

\bibitem[Plunian \& Alboussi\`ere(2020)]{Plunian2020}
{\sc \au{Plunian, F.} \& \au{Alboussi\`ere, T.}} \yr{2020}  \at{Axisymmetric
  dynamo action is possible with anisotropic conductivity}.  \jt{Physical
  Review Research}  \bvol{2},  \pg{013321}.

\bibitem[Plunian \& Alboussière(2021)]{Plunian2021}
{\sc \au{Plunian, F.} \& \au{Alboussière, T.}} \yr{2021}  \at{Axisymmetric
  dynamo action produced by differential rotation, with anisotropic electrical
  conductivity and anisotropic magnetic permeability}.  \jt{Journal of Plasma
  Physics}  \bvol{87}~(1),  \pg{905870110}.

\bibitem[Ponomarenko(1973)]{Ponomarenko1973}
{\sc \au{Ponomarenko, Y.B.}} \yr{1973}  \at{On the theory of hydromagnetic
  dynamos}.  \jt{Zh. Prikl. Mekh. \& Tekh. Fiz. (USSR)}  \bvol{6},
  \pg{47--51}.

\bibitem[Rincon(2019)]{Rincon2019}
{\sc \au{Rincon, F.}} \yr{2019}  \at{Dynamo theories}.  \jt{Journal of Plasma
  Physics}  \bvol{85}~(4),  \pg{205850401}.

\bibitem[Roberts(1972)]{Roberts1972}
{\sc \au{Roberts, G.O.}} \yr{1972}  \at{Dynamo action of fluid motions with
  two-dimensional periodicity}.  \jt{Philosophical Transactions of the Royal
  Society of London. Series A, Mathematical and Physical Sciences}  \bvol{271},
   \pg{411--454}.

\bibitem[Ruderman \& Ruzmaikin(1984)]{Ruderman1984}
{\sc \au{Ruderman, M.~S.} \& \au{Ruzmaikin, A.~A.}} \yr{1984}  \at{Magnetic
  field generation in an anisotropically conducting fluid}.  \jt{Geophysical \&
  Astrophysical Fluid Dynamics}  \bvol{28}~(1),  \pg{77--88}.

\bibitem[Ruzmaikin {\em et~al.\/}(1988)Ruzmaikin, Sokoloff \&
  Shukurov]{Ruzmaikin1988}
{\sc \au{Ruzmaikin, A.}, \au{Sokoloff, D.} \& \au{Shukurov, A.}} \yr{1988}
  \at{Hydromagnetic screw dynamo}.  \jt{Journal of Fluid Mechanics}
  \bvol{197},  \pg{39–56}.

\bibitem[Soward(1987)]{Soward1987}
{\sc \au{Soward, A.~M.}} \yr{1987}  \at{Fast dynamo action in a steady flow}.
  \jt{Journal of Fluid Mechanics}  \bvol{180},  \pg{267–295}.

\bibitem[Soward(1994)]{Soward1994}
{\sc \au{Soward, A.~M.}} \yr{1994}  \at{Fast dynamos} In \textit{Lectures on
  Solar and Planetary Dynamos} (ed. M. R. E. Proctor, A. D. Gilbert),181--217,
  Camb. Univ. Press.

\bibitem[Tobias(2021)]{Tobias2021}
{\sc \au{Tobias, S.M.}} \yr{2021}  \at{The turbulent dynamo}.  \jt{Journal of
  Fluid Mechanics}  \bvol{912},  \pg{P1}.

\bibitem[Vainshtein \& Zel'dovich(1972)]{Vainshtein1972}
{\sc \au{Vainshtein, S.I.} \& \au{Zel'dovich, Ya.~B.}} \yr{1972}  \at{Origin of
  magnetic fields in astrophysics}.  \jt{Usp. Fiz. Nauk}  \bvol{106},
  \pg{431--457}, [English transl.: Sov. Phys. Usp., Vol. 15, p. 159-172, 1972].

\bibitem[Zel'dovich(1957)]{Zeldovich1957}
{\sc \au{Zel'dovich, Ya.~B.}} \yr{1957}  \at{The magnetic field in the
  two-dimensional motion of a conducting turbulent liquid}.  \jt{Journal of
  Experimental and Theoretical Physics}  \bvol{4},  \pg{460}, [Russian original
  - ZhETF, Vol. 31, No. 3, p. 154, 1957].

\end{thebibliography}

\appendix
\section{Derivation of (\ref{eq:gammaBrBteta}$\mathit{a,}\mathit{b}$)}
\label{sec: eq of Br and Bteta}
The induction equation (\ref{eq:induction equation}) is derived from (\ref{eq:Faraday}) and (\ref{eq:Ohm}), such that
\begin{equation}
\partial_t \bB= \nabla \times (\bU \times \bB) - \nabla \times  \lbrack\sigma_{ij}\rbrack^{-1}\bJ,
\label{eq:indBJ}
\end{equation}
with, from (\ref{eq:Maxwell1}) and (\ref{eq:Ampere}), 
\begin{equation}
\bJ= \nabla \times  \lbrack\mu_{ij}\rbrack^{-1}\bB,
\label{eq:JB}
\end{equation}
Assuming axisymmetry ($\partial_{\theta}=0$) and considering the solenoidality of $\bB$ given by (\ref{eq:Maxwell2}), the curl of the cross product of $\bU=r\Omega(r) \be_{\theta}$ by $\bB=\left(B_r, B_{\theta}, B_z \right)\exp(\gamma t + \ii k z)$ takes the form 
\begin{equation}
\nabla \times (\bU \times \bB)=r\Omega' B_r \be_{\theta},
\label{eq:curlUB}
\end{equation}
where, from now, the exponential term is dropped for convenience.
From the definition (\ref{eq:resistivity and inv permeability tensors}$\mathit{a}$) of $\lbrack\sigma_{ij}\rbrack^{-1}$, we have
\begin{equation}
\lbrack\sigma_{ij}\rbrack^{-1} \bJ=\sigma_{\perp}^{-1}\begin{pmatrix} (1+\sigma c^2)J_r+ \sigma cs J_{\theta}\\ \sigma cs J_r + (1+\sigma s^2)J_{\theta}\\ J_z \end{pmatrix}.
\label{eq:magnetic field}
\end{equation} 
Taking the curl leads to
\begin{equation}
\nabla \times \lbrack\sigma_{ij}\rbrack^{-1} \bJ = \sigma_{\perp}^{-1}\begin{pmatrix} -\ii k\left[\sigma cs J_r + (1+\sigma s^2)J_{\theta}\right] \\ 
\ii k^{-1}\left[D_k(J_r)+\sigma c^2 k^2 J_r+ \sigma cs k^2 J_{\theta}\right] \\ 
 \frac{1}{r}\partial_r\left(r\left[\sigma csJ_r+(1+\sigma s^2)J_{\theta}\right]\right) \end{pmatrix}, \label{eq:current density1}
\end{equation}
where, again, $D_{\nu}(X)=\nu^2X-\partial_r\left(\frac{1}{r}\partial_r(rX)\right)$, and with $J_z=\ii k^{-1}\frac{1}{r}\partial_r(rJ_r)$ coming from the solenoidality of the current density.  

Similarily, we find that
\begin{equation}
\bJ=\nabla \times  \lbrack\mu_{ij}\rbrack^{-1}\bB= \mu_{\perp}^{-1}\begin{pmatrix} -\ii k\left[ \mu cs B_r + (1+\mu s^2)B_{\theta}\right] \\ 
\ii k^{-1}\left[D_k(B_r)+\mu c^2k^2 B_r+\mu csk^2 B_{\theta}\right] \\ 
 \frac{1}{r}\partial_r\left(r\left[\mu csB_r +(1+\mu s^2)B_{\theta}\right]\right) \end{pmatrix}. \label{eq:current density2}
\end{equation}
Combining (\ref{eq:current density1}) and (\ref{eq:current density2}) leads to
\begin{equation}
\nabla \times \lbrack\sigma_{ij}\rbrack^{-1} \left[\nabla \times  \lbrack\mu_{ij}\rbrack^{-1}\bB\right]= (\sigma_{\perp}\mu_{\perp})^{-1}\begin{pmatrix} F \\ 
G \\ \ii k^{-1}\left(\frac{1}{r}\partial_r(rF)\right) \end{pmatrix},
\label{eq:FG}
\end{equation}
with 
\begin{eqnarray}
F&=&\mu c^2 k^2 B_r +(1+\sigma s^2) D_k(B_r) - k^2 cs (\sigma -\mu)B_{\theta},\\
G&=&- cs (\sigma -\mu) D_k(B_r) + \sigma k^2 c^2 B_{\theta} + (1+\mu s^2) D_k(B_{\theta}).\;\;\;\;\;\;\;
\end{eqnarray}
Combining (\ref{eq:indBJ}), (\ref{eq:JB}), (\ref{eq:curlUB}) and (\ref{eq:FG}),   leads to (\ref{eq:gammaBrBteta}$\mathit{a,}\mathit{b}$).

\section{Derivation of (\ref{eq:Dkk}$\mathit{a,}\mathit{b}$)}
\label{sec:derivation of DoD}
Rewriting  (\ref{eq:gammaBrBteta2}$\mathit{a,}\mathit{b}$) in the form
\begin{subequations}\label{app:gammaBrBteta}
\begin{align}
D_k(B_r) &=-\left[\tilde{\gamma} B_r + \mu c^2k^2  B_r + \sigma s^2  D_k(B_r) - cs(\sigma -\mu)k^2 B_{\theta} \right] \label{app:gammaBr}\\
D_k (B_{\theta})  &= - \left[ \tilde{\gamma} B_{\theta} + \sigma c^2k^2 B_{\theta} + \mu s^2D_k (B_{\theta}) - cs(\sigma -\mu) D_k(B_r) \right] \label{app:gammaBteta},
\end{align}
\end{subequations}
and considering the linear combination $c$(\ref{app:gammaBr})+$s$(\ref{app:gammaBteta}),
leads to 
\begin{equation}
(1+\mu s^2) D_k(c B_r + s B_{\theta})= - (\tilde{\gamma}+\mu c^2 k^2) (c B_r + s B_{\theta}).
\end{equation}
Then, using the identity
\begin{equation}
(1+\mu s^2) D_k(X)=(1+\mu s^2)D_{k_{\mu}}(X) - (\tilde{\gamma}+\mu c^2 k^2) X, \;\;\;\;\; \label{eq:mu X}
\end{equation}
we find that
\begin{equation}
D_{k_{\mu}}(c B_r + s B_{\theta})=0. \label{app:A5}
\end{equation}

In a similar way, considering the linear combination $s D_k$((\ref{app:gammaBr}))-$ck^2$(\ref{app:gammaBteta}), leads to
\begin{equation}
(1+\sigma s^2) D_k(s D_k(B_r) -ck^2 B_{\theta})= - (\tilde{\gamma}+\sigma c^2 k^2) (s D_k(B_r) -ck^2 B_{\theta}).
\end{equation}
Then, using the identity
\begin{equation}
(1+\sigma s^2) D_k(X)=(1+\sigma s^2)D_{k_{\sigma}}(X) - (\tilde{\gamma}+\sigma c^2 k^2) X, \;\;\;\;\; \label{eq:sigma X}
\end{equation}
we find that
\begin{equation}
D_{k_{\sigma}}(s D_k(B_r) -ck^2 B_{\theta})=0. \label{app:A6}
\end{equation}

\section{Derivation of (\ref{eq:Btetalambda})}
\label{sec:Derivation de Bteta}
To obtain $B_{\theta}$ from $B_r$ by replacing (\ref{eq:Brlambda}) in (\ref{eq:gammaBr3}), we need to calculate $D_{k_{\sigma}}(B_r)$, given by
\begin{eqnarray}
D_{k_{\sigma}}(B_r)=&&
\left\{
\begin{split}
r< 1,&\;\;\;\;\;-s \lambda_{\mu} \frac{D_{k_{\sigma}}(I_1(k_{\mu} r))}{I_1(k_{\mu})}&  \\
r> 1,&\;\;\;\;\;-s \lambda_{\mu} \frac{D_{k_{\sigma}}(K_1(k_{\mu} r))}{K_1(k_{\mu})}&
\end{split}
\right. . \label{app:DkBrlambda}
\end{eqnarray}
With the help of the identity (\ref{eq:identity}) we have
\begin{eqnarray}
D_{k_{\sigma}}(B_r)=&&
\left\{
\begin{split}
r< 1,&\;\;\;\;\;-s \lambda_{\mu} (k_{\sigma}^2-k_{\mu}^2)\frac{I_1(k_{\mu} r)}{I_1(k_{\mu})}&  \\
r> 1,&\;\;\;\;\;-s \lambda_{\mu} (k_{\sigma}^2-k_{\mu}^2)\frac{K_1(k_{\mu} r)}{K_1(k_{\mu})}&
\end{split}
\right. . \label{app:DkBrlambda2}
\end{eqnarray}
Then replacing $k_{\sigma}$ and $k_{\mu}$ by their expressions given in (\ref{eq:ksigma kmu}$\mathit{a,}\mathit{b}$) and after some additional algebra leads to (\ref{eq:Btetalambda}).

\section{Derivation of the boundary condition (\ref{eq:A'cont})}
\label{sec:derivation of BC}
To write the continuity of $\partial_r B_r$ at $r=1$ we first need to calculate the expression of $\partial_r B_r$ at any $r$.
From the following identities satisfied for any $\nu$,
\begin{eqnarray}
\partial_r \left(I_1(\nu r)\right)&=&\nu I_0(\nu r) -\frac{1}{r}I_1(\nu r),\label{eq:diffI1}\\
\partial_r \left(K_1(\nu r)\right)&=&-\nu K_0(\nu r) -\frac{1}{r}K_1(\nu r),\label{eq:diffK1}
\end{eqnarray}
the expression of  $\partial_r B_r$ is obtained by deriving (\ref{eq:Brlambda}),
\begin{eqnarray}
\partial_r B_r=&&
\left\{
\begin{split}
r\le 1,&\;\;\;\;\;-s \frac{\lambda_{\sigma}}{I_1(k_{\sigma})} \left[k_{\sigma} I_0(k_{\sigma} r) -\frac{1}{r}I_1(k_{\sigma} r)\right]&  \\
       &\;\;\;\;\; -s \frac{\lambda_{\mu}}{I_1(k_{\mu})} \left[k_{\mu} I_0(k_{\mu} r) -\frac{1}{r}I_1(k_{\mu} r)\right]&  \\
r\ge 1,&\;\;\;\;\;s \frac{\lambda_{\sigma}}{K_1(k_{\sigma})} \left[ k_{\sigma} K_0(k_{\sigma} r) +\frac{1}{r}K_1(k_{\sigma} r) \right]&\\
       &\;\;\;\;\;s\frac{\lambda_{\mu}}{K_1(k_{\mu})} \left[ k_{\mu} K_0(k_{\mu} r) +\frac{1}{r}K_1(k_{\mu} r) \right]&
\end{split}
\right. . \;\;\;\;\;
\label{app:dAdr}
\end{eqnarray}
Then, from (\ref{app:dAdr}) writing the continuity of $\partial_r B_r$ at $r=1$ leads to (\ref{eq:A'cont}).

\section{Derivation of the current density $\bJ$ }
\label{sec:derivation of j}
The current density $\bJ$ given in (\ref{eq:current density2}) can be written, in its renormalized form, as
\begin{equation}
\bJ= \begin{pmatrix} -\ii k \phi \\ 
\ii k^{-1}\left[D_k(B_r)+\mu c^2k^2 B_r+\mu csk^2 B_{\theta}\right] \\ 
 \frac{1}{r}\partial_r\left(r\phi\right) \end{pmatrix}. \label{app:current density}
\end{equation}
with $\phi=\mu cs B_r + (1+\mu s^2)B_{\theta}$.

Replacing $B_r$ and $B_{\theta}$ by their expressions (\ref{eq:Brlambda}) and (\ref{eq:Btetalambda}),
leads to
\begin{eqnarray}
\phi=&&
\left\{
\begin{split}
r< 1,&\;\;\;c \left[\lambda_{\sigma}\frac{I_1(k_{\sigma}r)}{I_1(k_{\sigma})}  + \frac{s^2 \tilde{\gamma}}{c^2k^2}\lambda_{\mu}\frac{I_1(k_{\mu}r)}{I_1(k_{\mu})}\right]&  \\
r> 1,&\;\;\;c \left[\lambda_{\sigma}\frac{K_1(k_{\sigma}r)}{K_1(k_{\sigma})}  + \frac{s^2 \tilde{\gamma}}{c^2k^2}\lambda_{\mu}\frac{K_1(k_{\mu}r)}{K_1(k_{\mu})}\right] &
\end{split}
\right.,
\end{eqnarray}
and therefore to $J_r$.

Using the relations (\ref{eq:diffI1}) and (\ref{eq:diffK1}) written in the form
\begin{eqnarray}
\frac{1}{r}\partial_r \left(rI_1(\nu r)\right)&=&\nu I_0(\nu r) ,\label{eq:diffI2}\\
\frac{1}{r}\partial_r \left(rK_1(\nu r)\right)&=&-\nu K_0(\nu r) ,\label{eq:diffK2}
\end{eqnarray}
leads to 
\begin{eqnarray}
\frac{1}{r}\partial_r\left(r\phi \right)=&&
\left\{
\begin{split}
r< 1,&\quad\;\;\;\;c\left[\lambda_{\sigma}k_{\sigma}\frac{I_0(k_{\sigma}r)}{I_1(k_{\sigma})} + \frac{s^2 \tilde{\gamma}}{c^2k^2}\lambda_{\mu}k_{\mu}\frac{I_0(k_{\mu}r)}{I_1(k_{\mu})} \right]&   \\
r> 1,&\;\;\;-c\left[\lambda_{\sigma}k_{\sigma}\frac{K_0(k_{\sigma}r)}{K_1(k_{\sigma})} + \frac{s^2 \tilde{\gamma}}{c^2k^2}\lambda_{\mu}k_{\mu}\frac{K_0(k_{\mu}r)}{K_1(k_{\mu})} \right]& 
\end{split}
\right. ,
\end{eqnarray}
and therefore to $J_z$.

Using (\ref{eq:mu X}), we find that 
\begin{equation}
D_k(B_r)+\mu c^2 k^2B_r+\mu cs k^2 B_{\theta} = D_{k_{\mu}}(B_r)+\frac{\mu csk^2\phi-\tilde{\gamma}B_r}{1+\mu s^2}.
\end{equation}
From the expression of $B_r$ given by (\ref{eq:Brlambda}), we have
\begin{eqnarray}
D_{k_{\mu}}(B_r)=&&
\left\{
\begin{split}
r< 1,&\;\;\;-\frac{s\lambda_{\sigma}}{I_1(k_{\sigma})} D_{k_{\mu}}(I_1(k_{\sigma} r)) &  \\
r> 1,&\;\;\;-\frac{s\lambda_{\sigma}}{K_1(k_{\sigma})} D_{k_{\mu}}(K_1(k_{\sigma} r)) &
\end{split}
\right. . 
\end{eqnarray}
Using (\ref{eq:identity}) leads to
\begin{eqnarray}
D_{k_{\mu}}(B_r)=&&
\left\{
\begin{split}
r< 1,&\;\;\;-\frac{s\lambda_{\sigma}}{I_1(k_{\sigma})} (k_{\mu}^2-k_{\sigma}^2)I_1(k_{\sigma} r) &  \\
r> 1,&\;\;\;-\frac{s\lambda_{\sigma}}{K_1(k_{\sigma})} (k_{\mu}^2-k_{\sigma}^2)K_1(k_{\sigma} r) &
\end{split}
\right. , 
\end{eqnarray}
where we used the identity $D_{\nu}(I_1(k_{\nu} r))=D_{\nu}(K_1(k_{\nu} r))=0$.
After some algebra we find that
\begin{eqnarray}
D_{k_{\mu}}(B_r)+\frac{\mu csk^2\phi-\tilde{\gamma}B_r}{1+\mu s^2}=&&
\left\{
\begin{split}
r< 1,&\;\;\;\lambda_{\sigma}sk^2 \left(\frac{\sigma c^2+\tilde{\gamma}/k^2}{1+\sigma s^2}\right)\frac{I_1(k_{\sigma}r)}{I_1(k_{\sigma})}+s\tilde{\gamma}\lambda_{\mu}\frac{I_1(k_{\mu}r)}{I_1(k_{\mu})}&  \\
r> 1,&\;\;\;\lambda_{\sigma}sk^2 \left(\frac{\sigma c^2+\tilde{\gamma}/k^2}{1+\sigma s^2}\right)\frac{K_1(k_{\sigma}r)}{K_1(k_{\sigma})}+s\tilde{\gamma}\lambda_{\mu}\frac{K_1(k_{\mu}r)}{K_1(k_{\mu})} &
\end{split}
\right., 
\end{eqnarray}
leading to $J_{\theta}$.

Then the current density takes the following form

for $r<1$,
\begin{eqnarray}
J_r&=&-\ii k c \left[\lambda_{\sigma}\frac{I_1(k_{\sigma}r)}{I_1(k_{\sigma})}  + \frac{s^2 \tilde{\gamma}}{c^2k^2}\lambda_{\mu}\frac{I_1(k_{\mu}r)}{I_1(k_{\mu})}\right], \label{eq:Jrinf1}\\
J_{\theta}&=& \ii k^{-1} \left[\lambda_{\sigma}sk^2 \left(\frac{\sigma c^2+\tilde{\gamma}/k^2}{1+\sigma s^2}\right)\frac{I_1(k_{\sigma}r)}{I_1(k_{\sigma})}+s\tilde{\gamma}\lambda_{\mu}\frac{I_1(k_{\mu}r)}{I_1(k_{\mu})}\right],\label{eq:Jtetainf1}\\
J_z&=&c\left[\lambda_{\sigma}k_{\sigma}\frac{I_0(k_{\sigma}r)}{I_1(k_{\sigma})} + \frac{s^2 \tilde{\gamma}}{c^2k^2}\lambda_{\mu}k_{\mu}\frac{I_0(k_{\mu}r)}{I_1(k_{\mu})} \right],
\end{eqnarray}

for $r>1$,
\begin{eqnarray}
J_r&=&-\ii k c  \left[\lambda_{\sigma}\frac{K_1(k_{\sigma}r)}{K_1(k_{\sigma})}  + \frac{s^2 \tilde{\gamma}}{c^2k^2}\lambda_{\mu}\frac{K_1(k_{\mu}r)}{K_1(k_{\mu})}\right],\label{eq:Jrsup1}\\
J_{\theta}&=& \ii k^{-1} \left[\lambda_{\sigma}sk^2 \left(\frac{\sigma c^2+\tilde{\gamma}/k^2}{1+\sigma s^2}\right)\frac{K_1(k_{\sigma}r)}{K_1(k_{\sigma})}+s\tilde{\gamma}\lambda_{\mu}\frac{K_1(k_{\mu}r)}{K_1(k_{\mu})}\right],\label{eq:Jtetasup1}\\
J_z&=&-c\left[\lambda_{\sigma}k_{\sigma}\frac{K_0(k_{\sigma}r)}{K_1(k_{\sigma})} + \frac{s^2 \tilde{\gamma}}{c^2k^2}\lambda_{\mu}k_{\mu}\frac{K_0(k_{\mu}r)}{K_1(k_{\mu})} \right]. \label{eq:Jzsup1}
\end{eqnarray}

\section{Derivation of $Bz$ }
\label{sec:derivation of Bz}
From (\ref{eq:divB2}) we have
\begin{equation}
B_z = \ii k^{-1} (\frac{B_r}{r}+\partial_rB_r).
\label{eq:divB3}
\end{equation}
Then replacing (\ref{eq:Brlambda}) and (\ref{app:dAdr}) in (\ref{eq:divB3}), leads to
\begin{eqnarray}
B_z=&& \ii k^{-1} s
\left\{
\begin{split}
r< 1,&\;\;\;-\left[\lambda_{\sigma}k_{\sigma}\frac{I_0(k_{\sigma}r)}{I_1(k_{\sigma})} + \lambda_{\mu}k_{\mu}\frac{I_0(k_{\mu}r)}{I_1(k_{\mu})} \right]&   \\
r> 1,&\quad\quad\lambda_{\sigma}k_{\sigma}\frac{K_0(k_{\sigma}r)}{K_1(k_{\sigma})} + \lambda_{\mu}k_{\mu}\frac{K_0(k_{\mu}r)}{K_1(k_{\mu})}& 
\end{split}
\right. .
\label{eq:Bzlambda}
\end{eqnarray}

\end{document}